\documentclass[runningheads]{llncs}
\usepackage[T1]{fontenc}
\usepackage{tabularx} 
\usepackage{array}
\newcolumntype{S}{>{\arraybackslash}m{1.5cm}}
\newcolumntype{F}{>{\arraybackslash}m{1.5cm}}
\usepackage[table]{xcolor}

\usepackage{booktabs}
\usepackage{multirow}
\usepackage{hyperref}
\usepackage{tcolorbox}
\usepackage{color}
\usepackage{amsmath}
\usepackage{longtable}

\usepackage{graphicx}

\usepackage{soul} 
\usepackage{enumitem}
\usepackage[export]{adjustbox}
\usepackage{listings}
\usepackage{caption}
\usepackage{subcaption}
\usepackage{orcidlink}

\begin{document}



\title{LongEval at CLEF 2025: Longitudinal Evaluation of IR Systems on Web and Scientific Data}
\titlerunning{LongEval Lab at CLEF 2025}

\author{
Matteo Cancellieri\inst{1} \orcidlink{0000-0002-9558-9772} \and 
Alaa El-Ebshihy\inst{2,3}\orcidlink{0000-0001-6644-2360} \and
Tobias Fink\inst{2,3}\orcidlink{0000-0002-1045-8352} \and 
Maik Fr{\"o}be\inst{4}\orcidlink{0000-0002-1003-981X} \and  \\
Petra Galu\v{s}\v{c}\'{a}kov\'{a}\inst{5} \orcidlink{0000-0001-6328-7131} \and
Gabriela Gonzalez-Saez\inst{6}\orcidlink{0000-0003-0878-5263} \and
Lorraine Goeuriot\inst{6}\orcidlink{0000-0001-7491-1980} \and \\
David Iommi\inst{2}\orcidlink{0000-0002-4270-5709}\and
J\"{u}ri Keller\inst{7} \orcidlink{0000-0002-9392-8646} \and
Petr Knoth\inst{1} \orcidlink{0000-0003-1161-7359} \and
Philippe Mulhem\inst{6} \orcidlink{0000-0002-3245-6462} \and \\
Florina Piroi\inst{2,3} \orcidlink{0000-0001-7584-6439} \and
David Pride\inst{1} \orcidlink{0000-0002-7162-7252} \and
Philipp Schaer\inst{7} \orcidlink{0000-0002-8817-4632}
}
\authorrunning{M. Cancellieri et al.}
%
\institute{
The Open University, Milton Keynes, UK\footnote{Authors ordered alphabetically}
\and
Research Studios Austria, Data Science Studio, Vienna, Austria
\and
TU Wien, Austria
\and
Friedrich-Schiller-Universit{\"a}t Jena
\and
University of Stavanger, Stavanger, Norway
\and
Univ. Grenoble Alpes, CNRS, Grenoble INP\footnote{Institute of Engineering Univ. Grenoble Alpes.}, LIG, Grenoble, France
\and
TH Köln - University of Applied Sciences, Cologne, Germany
}

\maketitle

\begin{abstract}
  The LongEval lab focuses on the evaluation of information retrieval systems over time. Two datasets are provided that capture evolving search scenarios with changing documents, queries, and relevance assessments. Systems are assessed from a temporal perspective—that is, evaluating retrieval effectiveness as the data they operate on changes. In its third edition, LongEval featured two retrieval tasks: one in the area of ad-hoc web retrieval, and another focusing on scientific article retrieval. We present an overview of this year’s tasks and datasets, as well as the participating systems. A total of 19 teams submitted their approaches, which we evaluated using nDCG and a variety of measures that quantify changes in retrieval effectiveness over time.
\end{abstract}


\section{Introduction}

Information Retrieval (IR) systems are constantly challenged by the evolving search setting~\cite{DBLP:conf/sigir/Dumais14}. Document collections evolve, user information needs shift, and relevance judgments may vary over time~\cite{DBLP:conf/wsdm/AdarTDE09,ROBERTS2021103865,DBLP:conf/sigir/TikhonovBBOKG13}. These temporal dynamics have strong implications to the long-term effectiveness of retrieval models. It is known that search is sensitive to temporal shifts~\cite{DBLP:journals/ftir/KanhabuaBN15,liu2024robustneuralinformationretrieval} and that incorporating historical signals can enhance retrieval robustness~\cite{DBLP:conf/clef/AlexanderFHSHHP24,keller2024leveraging}. Additionally, in modern IR, the systems are updated or retrained often, making them a dynamic component themselves in the evolving search setting. 

While these temporal factors strongly influence retrieval effectiveness, they are often overlooked in standard IR evaluation protocols, which typically assume a static test collection. The advantages of such datasets are that they are easily used to evaluate and test systems. Some data sets, like CORD19, contain documents collected at different points in time, showing differences in the set of documents from one collection time to another. We have shown previously that the 
the ranking of systems varies over time and that the most effective system is not necessarily also the system that performs the most consistently~\cite{longevaloverview2023,longevalCLEF2024overview,keller2024evaluation}. This shows how the experimental setup strongly influences the measured effectiveness.

With the aim of tackling this challenge of making models have persistent quality over time, the objective of the LongEval lab is twofold: (i) to systematically assess how performance of retrieval systems changes over time as test collections evolve, and (ii) to propose improved methods that mitigate performance drop by making models more robust over time.

The third edition of the LongEval lab~\cite{longevalecir2025} was part of the Conference and Labs of the Evaluation Forum (CLEF) 2025. In this edition it consisted of two retrieval tasks: Task 1 - WebRetrieval, which is the classical web case, and Task 2 - SciRetrieval, which is for scientific search. Task 1 evaluates retrieval robustness over time using an evolving web search collection, while Task 2 follows a similar setup but uses scholarly publications as the underlying document corpus.

\section{Tasks Description}

In contrast to traditional IR evaluation that rely on one static datasets, LongEval\,2025 explores the effect of changing datasets on the retrieval systems and measured effectiveness.
Similarly to the LongEval\,2023 and 2024 Retrieval Tasks~\cite{longevaloverview2023,longevalCLEF2024overview}, we focus on a setup in which the datasets are evolving. In concrete terms, differences between datasets can be the addition, removal, or the change of documents and queries. Each evolved state of a dataset is captured as a new snapshot, which forms the basis for new experiments. We evaluate systems by computing efficiency metrics on the experimental approaches (or runs) submitted by the participants who designed those systems. The two main scenarios considered in our evaluation focus on single systems and multiple systems:

\noindent\textbf{A single-system in evolving setup:} Each system is trained once, on a collection snapshot at a given timestamp, and evaluated on later collection snapshots. This setup assesses how well a system maintains retrieval effectiveness over time without retraining.

\noindent\textbf{Multi-system comparison:} Each system is compared to the other systems, across several snapshots. This evaluation setup is to provide more information about systems' stability and robustness compared to each-other.

\subsection{Task 1: WebRetrieval}
\label{sec:webretrieval}

The WebRetrieval task investigates how IR systems cope with evolving web search collections. The dataset consists of a series of monthly snapshots between June 2022 to August 2023, extracted from the French search engine Qwant\footnote{Qwant search engine: \url{https://www.qwant.com/}}, each containing documents and corresponding user queries. 

\subsection{Task 2: SciRetrieval}
\label{sec:sciretrieval}
The SciRetrieval task extends the LongEval Lab to the domain of academic search. It aims to evaluate the temporal robustness of IR systems when retrieving scholarly publications from an evolving corpus. The dataset is derived from the CORE collection\footnote{CORE search engine: \url{https://core.ac.uk/}}, which aggregates open access research outputs from repositories and journals worldwide. As of this edition, CORE is one of the largest platforms of openly available full-text scholarly documents.
\section{Datasets}
\label{sec:datasets}
This section describes the datasets used in the two retrieval tasks, including data sources, snapshot construction, document and query statistics, and relevance judgments methods. All datasets are available from the TU Wien Research Data Repository~\cite{data_web_2025,data_sci_2025_test,data_sic_2025_train} or through the LongEval IR\_datasets integration~\cite{ir_datasets_longeval,DBLP:conf/sigir/MacAvaneyYFDCG21}\footnote{GitHub: \url{https://github.com/clef-longeval/ir-datasets-longeval}}.

\subsection{WebRetrieval Dataset}
The dataset for this task was provided by the French search engine Qwant. It consist of the queries issued by the users of this search engine, cleaned Web documents, which were 1) selected to correspond to the queries, and 2) to add additional noise, and relevance judgments, which were created using a click model. The dataset as of 2022 is fully described in~\cite{2023longevalretrieval}. Later additions to the LongEval WebRetrieval collection have followed the same collection procedure.

The 2025 dataset includes all data from the 2023 and 2024 editions, along with newly added, previously unreleased months. The training dataset consists of 19 million French documents (June 2022 - February 2023) and 119,341 queries with computed relevance assessments based on a simplified Dynamic Bayesian Network (sDBN) Click Model~\cite{Chapelle2009-rg,Chuklin2015}, acquired from real users of the French Qwant search engine. Compared to the previous datasets, the data were processed differently to combine similar queries and unify IDs. Therefore, direct comparisons are not possible, although all datasets have overlapping snapshots.

The test collection spans 7 months of data (March 2023 - August 2023) and consists of 14 million documents and 63,416 queries. Each month is captured in one snapshot. They are similar in structure as the training snapshot, except that they do not contain any relevance assessments. Participants submitted their runs for each snapshot, using the same system trained only on the training dataset. 
The total data for this task consists of 33 million documents and 182,757 queries, provided by Qwant.

\begin{figure}[ht]
\centering
  \includegraphics[width=1\linewidth]{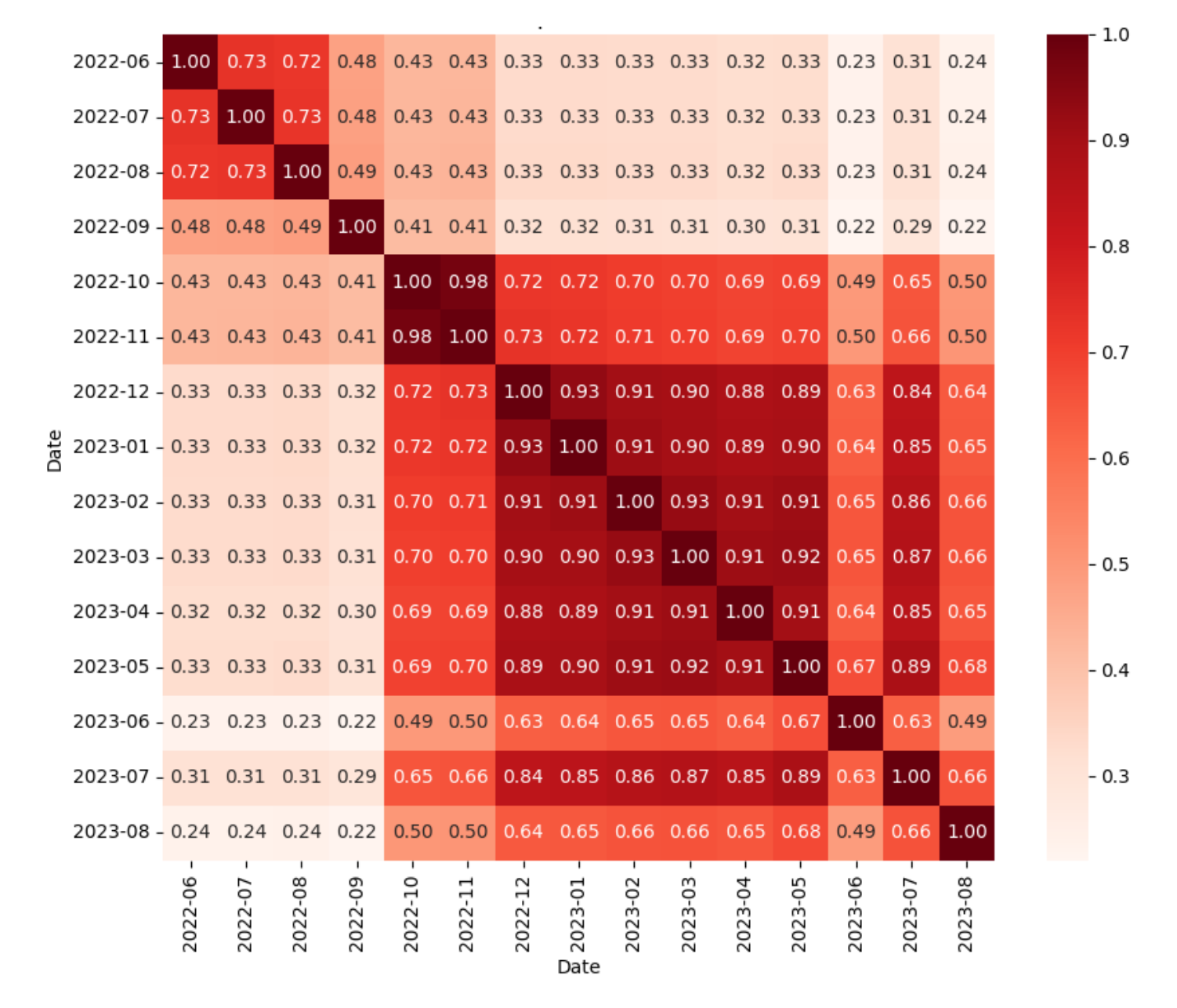}
  \caption{Document overlap between monthly snapshots of the WebRetrieval dataset. Lighter colors indicate lower overlap between collections.}
  \label{fig:webretrieval_docs_overlap}
\end{figure}

Figure~\ref{fig:webretrieval_docs_overlap} shows the document overlap between each pair of monthly snapshots in the WebRetrieval dataset. The values reflect the proportion of shared documents across snapshots. As shown, overlap is highest between consecutive months and decreases over time. More specifically, earlier snapshots (e.g., mid-2022) share substantially fewer documents with later ones (e.g., mid to late 2023), demonstrating the evolving nature of the dataset. This highlights the challenge posed to the retrieval models, which should maintain effectiveness despite the changes of the collection over time.

\subsection{SciRetrieval Dataset}
The data for this task were acquired from the CORE\footnote{CORE (COnnecting REpositories) \url{https://core.ac.uk/}} collection of scholarly documents. The dataset includes cleaned user queries and documents returned as results to these queries, along with additional negative, randomly selected scientific articles. Similar to the WebRetrieval task, the relevance judgments were created using a click model~\cite{Chapelle2009-rg,Chuklin2015}, based on click signals from user logs (e.g., PDF downloads).

As this is the first iteration of the SciRetrieval task, the dataset includes only two snapshots:
(1) The first snapshot spans mid-November 2024 to mid-December 2024, and
(2) The second snapshot was collected in January 2025.

The training dataset (from the first snapshot) includes approximately 2 million documents, 393 queries, and 4,262 relevance assessments. The test dataset includes over 1 million documents, 492 new queries (from the second snapshot), and 99 held-out queries selected from the first snapshot. Like in the WebRetrieval task, the test collections mirror the structure of the training data but do not contain relevance assessments. Participants submitted their runs for both test sets using a system trained only on the training dataset.
The total data for this task consists of over 3 million documents and 984 queries.

\section{Submissions}
\label{sec:submissions}
We received 45 runs to the WebRetrieval task and 23 runs to the SciRetrieval task from 19 participating teams. Of these 19 teams 12 submitted a notebook paper. 
The submissions were done through the TIRA platform~\cite{froebe:2023b} where participants could upload either runs, software, or both. 36 further teams registered in tira but never submitted any run. We collected the descriptions and metadata of each run in the ir\_metadata specification~\cite{breuer:2022}. Table~\ref{tab:ir_metadata} gives an overview of the ir\_metadata of the submitted approaches.

In the following, we give a brief summary of the submissions based on the information provided by the participants.

\begin{table}[t]
    \centering
    \setlength{\tabcolsep}{6pt}
    \caption{Teams participating to the LongEval 2025 lab.}
   \begin{tabular}{l c l p{4.5cm} c}
    \toprule
    Team     & Team size & Task & Location 
    & Notebook  
    \\
     & (\# persons) & & 
     &
     \\
    \midrule    
    3DS2A & 6 & Web & Padua, Italy  
    & \cite{3DS2A} 
    \\
    BASETTE & 4 & Web & Padua, Italy 
    & \cite{BASETTE}
    \\
    CIR & 19 & Web & Cologne, Germany 
    & \cite{CIR} 
    \\ 
    CIR\_cluster & 3 & Web & Cologne, Germany 
    & \cite{CIRcluster} 
    \\
    DataHunter & 5 & Web & Padua, Italy 
    & \cite{DataHunter}
    \\
    DS@GT & 3 & Web & Georgia, USA 
    & \cite{DS@GT}
    \\
    EAIiIB & 4 & Sci & Cracow, Poland 
    & \cite{EAIiIB}
    \\
    OpenWebSearch & 7 & Sci & Nijmegen, The Netherlands; Jena and Kassel, Germany 
    & \cite{OWS}
    \\
    RACOON & 5 & Web & Padua, Italy 
    & \cite{RACOON}
    \\
    RAND & 7 & Web & Padua, Italy 
    & \cite{RAND}
    \\    RISE & 7 & Web & Padua, Italy 
    & \cite{RISE}
    \\
    SARD    & 5 & Web & Padua, Italy 
    & \cite{SARD} 
    \\
    
    \bottomrule
    \end{tabular}    
    \label{tab:submissions}
\end{table}

\begin{table}[t]
    \centering
    \setlength{\tabcolsep}{6pt}
    \renewcommand{\arraystretch}{1.25}
    \caption{Overview of the collected ir\_metadata according to (a) frequently used software libraries, (b) implemented retrieval paradigms, and (c) available git repositories.}
    \label{tab:ir_metadata}

    \begin{minipage}[t]{0.45\textwidth}
    \begin{tabular}{@{}l c@{}}
    \multicolumn{2}{l}{(a) Most popular libraries}\\
    \toprule
    Library     & Teams \\
    \midrule    
    python-terrier & 7 \\
    numpy & 4 \\
    SQLAlchemy & 4 \\
    scikit-learn & 4 \\
    pandas & 4 \\
    transformers & 3 \\
    torch & 3 \\
    ir-datasets-longeval & 3 \\
    Lucene & 3 \\
    Yaml & 2 \\
    \bottomrule
    \end{tabular}
    \end{minipage}%
    \begin{minipage}[t]{0.4\textwidth}
    \renewcommand{\arraystretch}{1}
    \begin{tabular}{@{}l c c@{}}
    \multicolumn{3}{l}{(b) Retrieval paradigms}\\
    \toprule
    \textbf{Paradigm}     &  \multicolumn{2}{c}{\textbf{Used}} \\
    \cmidrule(l){2-3}
    & Yes & No\\
    \midrule    
    Lexical & 84 & 23 \\
    Deep Neural & 60 & 47 \\
    Sparse Neural & 5 & 102 \\
    Dense Neural & 59 & 48 \\
    Single Stage & 67 & 40 \\
    \bottomrule
    \vspace*{2cm}
    \end{tabular}
    \end{minipage}%

    \begin{minipage}[t]{0.49\textwidth}
    \phantom{foo}
    \end{minipage} %
    \begin{minipage}[t]{0.45\textwidth}
    \vspace*{-2.6cm}
    \begin{tabular}{@{}l c c@{}}
    \multicolumn{3}{l}{(c) Repositories}\\
    \toprule
    \textbf{Hoster}     &  \textbf{Public}  &  \textbf{Private} \\
    \midrule    
    Bitbucket & 0 & 8 \\
    Github & 10 & 4 \\
    \bottomrule
    \end{tabular}
    \end{minipage}
    
\end{table}
\vspace{-0.2cm}
\paragraph{3DS2A~\cite{3DS2A}:} 
Besides a traditional BM25 search pipeline, the system by Team 3DS2A utilizes more sophisticated techniques like chunk-based search, where documents are divided into semantically coherent text segments - called chunks - before indexing. The effectiveness of chunk-based approaches has been highlighted in both neural and classical IR frameworks. Additionally, pseudo-relevance feedback and reranking based on sentence embeddings were used to enhance the retrieval capability of the queries. The system was based on Lucene.
\vspace{-0.2cm}
\paragraph{BASETTE~\cite{BASETTE}:}
Team BASETTE's main priority was optimizing performance on limited consumer hardware,
deliberately avoiding the use of GPUs or other specialized computational resources. The system relies on classical IR techniques and is designed to run both indexing and retrieval in a multithreaded fashion to ensure high execution speed. One example of this was to process the indexing in-memory and never write it to disk. Parameters of the search system were optimised using Optuna. The team chose to discard a bunch of approaches due to limited hardware resources or integration complexity, like WordNet term expansion, the integration of the Duckling framework to extract temporal expressions from the documents, or neural ranking with CamemBERT.
\vspace{-0.2cm}
\paragraph{CIR~\cite{CIR}:} 
This multi-team submission summarized five different teams: CIR\_SchaeredRetrieval, CIR\_SuperTeam123, CIR\_Sauerkraut, CIR\_JMFT, and CIR\_fair\_schaer. The groups had five different approaches and motivations to test in the LongEval setting:  (1) Finding time-dependent queries with the help of LLMs and to treat these queries differently by boosting their retrieval scores based on the categorization; (2) Finding time-dependent queries and scoring them on a scale from 0 to 1 and to use that score to influence the final ranking; (3) Using relevance information from older sub-collections and to use relevance feedback on the current sub-collection by using query expansion using tf-idf; (4) Boosting known relevant documents-query pairs from older sub-collections but comparing the similarity of old and recent documents; Finally, (5) a neural relevance re-ranking based on a topcial semantic clustering. All systems used PyTerrier’s BM25 as the foundational retrieval system. 
\vspace{-0.2cm}
\paragraph{CIR\_cluster~\cite{CIRcluster}:} 
The submission by Team CIR\_cluster aims to leverage historical information,  such as past relevance judgments. In a previous submission, this information was used in two different ways: Query Boost and Relevance Feedback. Both methods are limited when no prior information is available, which is the case when queries can not be mapped due to slight variations. It was observed that many similar queries are captured in the test collections. Often, they even differ only on a lexical level, such as spelling or word order. Therefore, Team CIR\_cluster aimed to identify query variants – queries that relate to the same information need but express it in different ways, as documents relevant to a given query might also be relevant to its semantic variants. Based on this assumption, they cluster queries and link previously unseen queries to the history of their query variants. 
\vspace{-0.2cm}
\paragraph{DataHunter~\cite{DataHunter}:}
Team DataHunter compared a traditional BM25 query-based searcher with a neural reranking system. The latter used an inverse square rank fusion based on BM25 with RM3 expansion, SPLADE (sparse transformer-based retrieval), and a cross-encoder CamemBERT re-ranker (top-100 reranking). Next to the retrieval system, the team also conducted a statistical analysis using ANOVA and Tukey HSD tests to determine whether performance differences are significant over time, providing insights into the robustness and generalizability of IR models in dynamic environments.
\vspace{-0.2cm}
\paragraph{DS@GT~\cite{DS@GT}:}
Team DS@GT developed an IR pipeline to observe temporal variance using topic modeling and a two-phase retrieval system involving query expansion based on the Gemini LLM. The workflow begins with a Parquet ingestion pipeline for the LDA topic modeling and sentence transformer processing. The transformed documents are stored in a shared directory where the Pyserini system was used to perform BM25 retrieval. When an input query is received, the system references a precomputed mapping index of Qwant search queries, each mapped to expanded queries generated using Gemini. 
\vspace{-0.2cm}
\paragraph{EAIiIB~\cite{EAIiIB}:}
The EAIiIB team participated in the Sci task and compared classical lexical retrieval, dense vector-based retrieval, and hybrid approaches, incorporating reranking via cross-encoders. They experimented with a reduced and the full data set and used the following models to build their baseline: BM25, MiniLM-L6-v2, E5-large-v2. Additionally, some hybrid pipelines were introduced, combining different approaches. In their setting, dense and cross-encoder reranking outperforms all lexical and hybrid configurations.
\vspace{-0.2cm}
\paragraph{OpenWebSearch~\cite{OWS}:}
Team OpenWebSearch used different web crawls of the CORE search engine to build their submissions on. They crawled the top-25 documents from CORE for quries based on different fields (title, abstract, and full text). They had two underlying assumptions: First, a practical search engine should make only incremental improvements to existing rankings, and second, that most relevant documents are already among the top results shown by the CORE engine, especially since LongEval defines relevance based on user clicks. Building on these initial rankings, different techniques to enhance retrieval effectiveness were tested: (1) qrel-boosting based on past relevance, RM3 keyquery expansion, cluster-based boosting, monoT5 re-ranking of top results, and user intent prediction. Each of these methods is designed to re-rank or selectively boost documents within the CORE results to increase the likelihood of retrieving relevant content. 
\vspace{-0.2cm}
\paragraph{RACOON~\cite{RACOON}:}
The search system developed by Team RACOON incorporates a French lemmatizer within the analyzer, a LLM-based Query Expansion (QE) module, a Relevance Feedback (RF) mechanism based on query assessment over the
preceding months, and an LLM-based Elo reranking strategy. The custom lexicon lemmatizer is based on the Lexique des formes fléchies du français (LEFFF), a large-scale morphological and syntactic lexicon for French. The QE pipeline was using LLaMA 3 70B and LLaMA 4 Scout 17B and instructed the models to (1) generate 20 semantically related expressions to a query, (2) provide a passage that answers the initial query, (3) listing synonyms, and (4) infer the user's intent explicitly. The RF was based on worked presented in 2024 \cite{keller2024leveraging}. The reranker was finally implemented using either an SBERT embedding cosine similarity score, an ms-marco-MiniLM-L6-v2 setting, or an ELO-pseudo ranking method, where two documents were randomly compared at a time and an LLM decided on their appropriateness to the query. 
\vspace{-0.2cm}
\paragraph{RAND~\cite{RAND}:}
Team RAND built upon a classic Lucene pipeline and tested various configurations, including different tokenizers, filters, stoplists, stemmers, etc. They also used the pre-trained sentence embedding model multi-qa-MiniLM-L6-cos-v1 and Llama-3.2-1B-Instruct to include a semantic scoring for re-ranking the initial Lucene results. Additionally, they included the Lucene ICUFoldingFilter, which applies search term folding to Unicode text, including accent removal and case folding, among other features. This proved to be useful, as the French language contains many words with diacritical marks (accents), which can impact search and term matching. 
\vspace{-0.2cm}
\paragraph{RISE~\cite{RISE}:}
The system of Team RISE incorporates a modular architecture, including a parser, an analyzer, an indexer, and a searcher. It also includes query translation and expansion using the Gemini LLM, and a non-neural reranking component to enhance retrieval quality. The emphasis was placed on optimizing indexing and search speed through multithreading, improving relevance by crafting a title for each document, and improving the content of the URL-based document based on the alignment between user queries and the document’s URL. The system extracted titles from documents and query terms from the URL. Both was not marked explicitly in the dataset to give the extracted titles a higher weight in the final ranking. 
\vspace{-0.2cm}
\paragraph{SARD~\cite{SARD}:} 
The SARD team focused on the development of custom language-specific analyzers for the preprocessing of documents and queries, and also the correction of user queries using GPT-4 Turbo. The use of GPT-4 Turbo aims to rewrite and clarify ill-formed or noisy queries without altering their intent, resulting in more precise and effective matching with the indexed content. The Basesystem is implemented on Lucene, and a language-detection process was introduced to differentiate the two main languages of the corpus. Additionally, two distinct query processes were employed: BooleanQuery and PhraseQuery. An LLM-based QE approach was discarded due to its high costs and processing time, and was replaced with a dictionary-based QE (WordNet for English and WOLF for French).
\section{Results}
\label{sec:results}
The submitted retrieval approaches are evaluated along different dimensions and measures, and different snapshots.
For the SciRetrieval task two test snapshots were available to compare the system effectiveness. For the WebRetrieval task, six test snapshots were available. In this paper we detail the evaluation for the WebRetrieval task for the following two cases: (a) short-term effectiveness change evaluation, comparing scores for the snapshots 2023-03 and 2023-05, and (b) long-term effectiveness change evaluation for scores on snapshots 2023-03 and 2023-08. 
\subsection{Effectiveness}
\begin{figure}[ht]
\centering
        \includegraphics[width=0.99\linewidth]{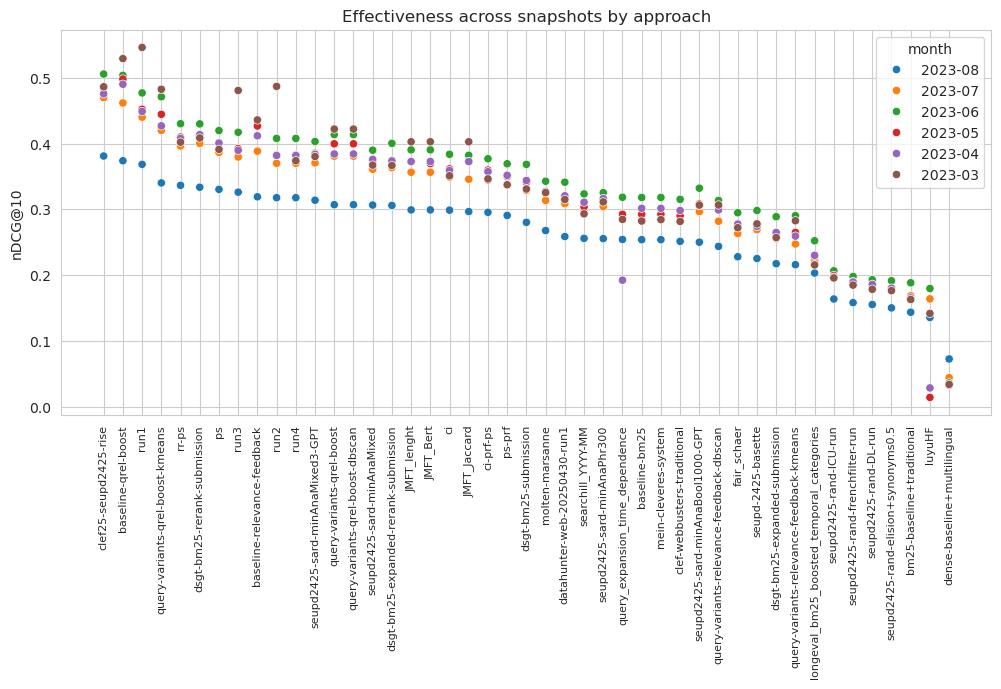}

        \caption{nDCG@10 for all approaches and test snapshots in the WebRetrieval task.}
      \label{fig:scatter-web} 
\end{figure}

\begin{figure}[h!]
\centering
    \begin{subfigure}[t]{0.49\textwidth}
        \centering
        \includegraphics[width=0.99\linewidth]{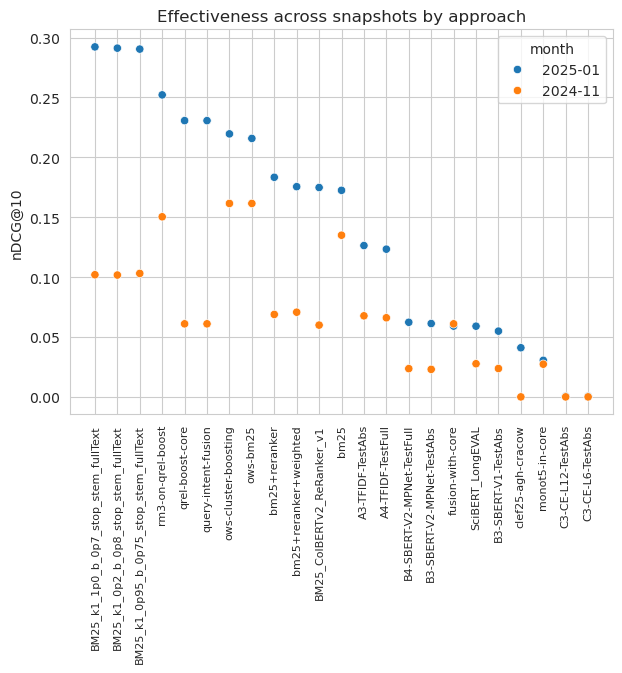}
        \caption{nDCG@10}
    \vspace{-0.2cm}
    \label{fig:scatter-sci}
    \end{subfigure}%
        \begin{subfigure}[t]{0.49\textwidth}
        \centering
    \includegraphics[width=0.99\linewidth]{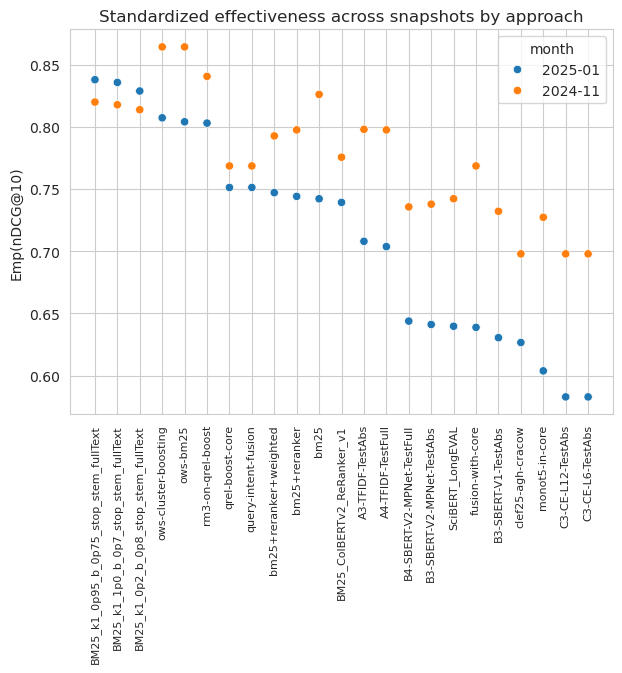}
        \caption{Standardized nDCG@10}
    
    \vspace{-0.2cm}
    \label{fig:scatter-sci-emp}
    \end{subfigure}%
    \caption{Effectiveness measured by nDCG@10 across all approaches and test snapshots of the SciRetrieval task.}
\end{figure}

The retrieval effectiveness is assessed by nDCG and nDCG@10. The results per snapshot and approach are presented in Tables~\ref{tab:web-results} for the Web task and Table~\ref{tab:sci-results} for the Sci task. The effectiveness measured by nDCG@10 for the test snapshots is also visualized in Figure~\ref{fig:scatter-web} for the Web task and in Figure~~\ref{fig:scatter-sci} for the Sci task. 
Especially for the Web task, nDCG@10 is considered to match the retrieval setting of Web search well, where users often focus on the first retrieval results. 
Some participants submitted multiple versions of the same approach that do not or only slightly differ. In these cases, only the best or newest approach was considered. For the Sci task, one team submitted only for the 2025-01 snapshot, and one team submitted runs with really few documents ranked.

\begin{figure}[h!]
\centering
        \includegraphics[width=0.99\linewidth]{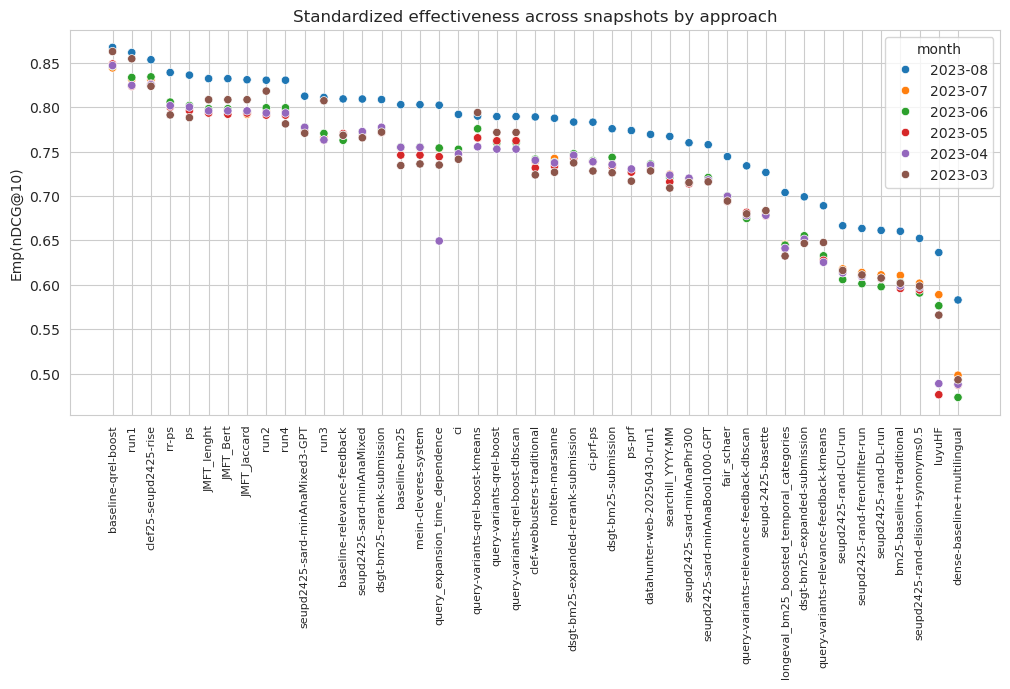}
    
      \caption{Standardized nDCG@10 for all approaches and test snapshots in the WebRetrieval task.}
      \vspace{-0.5cm}
      \label{fig:scatter-web-emp}
\end{figure}

The top three approaches for the Web task appear to be relatively stable over time. For the later snapshots, the approach \texttt{clef25-seupd2425-rise} from team RISE outperforms the baseline \texttt{qrel-boost}. These two approaches are followed by the \texttt{run1} of team RACOON, \texttt{rr-ps} of team 3DS2A, and \texttt{query-variants-qrel-boost-kmeans} of team CIR-CLUSTER as measured by nDCG@10. For nDCG@1000, this order slightly varies. Especially in 2023-03, the approaches \texttt{run1} and \texttt{run2} from team RACOON perform better. The gradient of both result tables recommends a relatively stable system ranking for both measures.

The top three approaches for the Sci task vary by the snapshot. For the first snapshot 2024-11 (within time), the approaches \texttt{ows-cluster-boosting} and \texttt{ows-bm25} are on par, followed by \texttt{rm3-on-qrel-boost}. All approaches were submitted by the Open Web Search team. Regarding the second snapshot 2025-01 (short term), the three approaches \texttt{BM25\_k1\_1p0\_b\_0p7\_stop\_stem\_fullText}, \texttt{BM25\_k1\_0p2\_b\_0p8\_stop\_stem\_fullText}, and \texttt{BM25\_k1\_0p95\_{\allowbreak}b\_0p75\_stop\_{\allowbreak}stem\_fullText} by the team Academy Retrievals dominate. As depicted by the gradient in Table~\ref{tab:sci-results}, the ranking of systems appears to be more stable across measures than time.

To compare the systems' effectiveness across snapshots, we also present standardized nDCG@10 scores using empirical standardization, as in \cite{urbano2019new} in Figure~\ref{fig:scatter-web-emp} for the Web task and  in Figure~~\ref{fig:scatter-sci-emp} for the Sci task. Each query of the snapshot is standardized according to the performance of all participating systems. Finally, a standardized nDCG@10 mean is computed.

When considering the standardized performance, the best systems in the Web task remain the same but in a different order (Figure~\ref{fig:scatter-web-emp}): the baseline \texttt{qrel-boost} achieves the highest standardized performance in snapshot 2023-08, with \texttt{run1} from team RACOON in second place. The baseline \texttt{qrel-boost} also stands out for its stability. Since it is built using historical data, we can hypothesize that this contributes to its temporal stability, as its performance variations reflect the differences between snapshots.
We observed similar results for the Sci task (Figure~\ref{fig:scatter-sci-emp}): the best systems in snapshot 2025-01 remain the same and have the most stable standardized performance. This stability suggests that their performance varies in line with the general trend of the ensemble of participating systems, as represented by the empirical standardization function.

\subsection{Ranking of Systems}
We inspect system rankings between two snapshots to better understand how the approaches evolve relative to each other over time. This evaluation should provide insights into how robust the system ranking is. As discussed before, the gradients of Table~\ref{tab:web-results} indicate a similar ranking of systems for the WebRetrieval task. Table~\ref{tab:sci-results} shows a less similar system ranking for the SciRetrieval case.

To further investigate this, we compare the system order between two snapshots based on the systems' ranks with Kendall's Tau and based on the measured effectiveness scores by Pearson’s correlation. All correlations are based on the nDCG@10 scores and rankings.

For the Web task, the short-term change (2023-03 to 2023-05) and long-term change (2023-03 to 2023-08) are compared. The Pearson correlation for the short-term change is 0.954, indicating an extremely strong positive linear relationship. The long-term change (2023-03 to 2023-08) shows an even slightly stronger correlation with 0.965. The Kendall also indicates a high correlation. For the short-term change, 0.930 was measured, and for the long-term, 0.893. All p-values indicate significance. 

For the Sci task, only two snapshots are available for comparison. The Person's correlation is much weaker compared to the Web correlations but is with 0.708 still strong. The Kendall’s tau of 0.500 indicates a moderate‐to‐strong positive. Like before, the p-values are significant. 

The ranking of systems shows a stronger correlation when standardized performances are compared. Table~\ref{tab:sci-results-std} presents the results for the scientific task, with a Pearson correlation of 0.848 and a Kendall's tau of 0.587.

\begin{table}
\caption{Evaluation results for the short-term and long-term changes of the WebRetrieval task. The results are sorted by nDCG@10 for the 2023-08 snapshot.}
\label{tab:web-results}
\resizebox{\textwidth}{!}{%
\begin{tabular}{lccc@{\hskip 5pt}ccc}
\toprule
 & \multicolumn{3}{c}{nDCG@10} &  \multicolumn{3}{c}{nDCG@1000} \\
Approach & 2023-03 & 2023-05 & 2023-08  & 2023-03 & 2023-05 & 2023-08 \\
 \midrule
clef25-seupd2425-rise \cite{RISE} & {\cellcolor[HTML]{00692A}} \color[HTML]{F1F1F1} 0.487 & {\cellcolor[HTML]{004D1F}} \color[HTML]{F1F1F1} 0.484 & {\cellcolor[HTML]{00441B}} \color[HTML]{F1F1F1} 0.381 & {\cellcolor[HTML]{005F26}} \color[HTML]{F1F1F1} 0.541 & {\cellcolor[HTML]{00441B}} \color[HTML]{F1F1F1} 0.543 & {\cellcolor[HTML]{00441B}} \color[HTML]{F1F1F1} 0.422 \\
baseline-qrel-boost \cite{CIRcluster} & {\cellcolor[HTML]{004E1F}} \color[HTML]{F1F1F1} 0.529 & {\cellcolor[HTML]{00441B}} \color[HTML]{F1F1F1} 0.498 & {\cellcolor[HTML]{004A1E}} \color[HTML]{F1F1F1} 0.374 & {\cellcolor[HTML]{005321}} \color[HTML]{F1F1F1} 0.561 & {\cellcolor[HTML]{00451C}} \color[HTML]{F1F1F1} 0.540 & {\cellcolor[HTML]{005221}} \color[HTML]{F1F1F1} 0.407 \\
run1 \cite{RACOON} & {\cellcolor[HTML]{00441B}} \color[HTML]{F1F1F1} 0.546 & {\cellcolor[HTML]{006328}} \color[HTML]{F1F1F1} 0.452 & {\cellcolor[HTML]{005120}} \color[HTML]{F1F1F1} 0.368 & {\cellcolor[HTML]{00441B}} \color[HTML]{F1F1F1} 0.588 & {\cellcolor[HTML]{005924}} \color[HTML]{F1F1F1} 0.508 & {\cellcolor[HTML]{005321}} \color[HTML]{F1F1F1} 0.405 \\
query-variants-qrel-boost-kmeans \cite{CIRcluster} & {\cellcolor[HTML]{006C2C}} \color[HTML]{F1F1F1} 0.483 & {\cellcolor[HTML]{00682A}} \color[HTML]{F1F1F1} 0.445 & {\cellcolor[HTML]{026F2E}} \color[HTML]{F1F1F1} 0.340 & {\cellcolor[HTML]{016E2D}} \color[HTML]{F1F1F1} 0.517 & {\cellcolor[HTML]{006428}} \color[HTML]{F1F1F1} 0.490 & {\cellcolor[HTML]{026F2E}} \color[HTML]{F1F1F1} 0.376 \\
rr-ps \cite{3DS2A} & {\cellcolor[HTML]{2A924A}} \color[HTML]{F1F1F1} 0.402 & {\cellcolor[HTML]{107A37}} \color[HTML]{F1F1F1} 0.410 & {\cellcolor[HTML]{05712F}} \color[HTML]{F1F1F1} 0.337 & {\cellcolor[HTML]{1D8640}} \color[HTML]{F1F1F1} 0.462 & {\cellcolor[HTML]{006D2C}} \color[HTML]{F1F1F1} 0.475 & {\cellcolor[HTML]{016E2D}} \color[HTML]{F1F1F1} 0.378 \\
dsgt-bm25-rerank-submission \cite{DS@GT} & {\cellcolor[HTML]{278F48}} \color[HTML]{F1F1F1} 0.409 & {\cellcolor[HTML]{0E7936}} \color[HTML]{F1F1F1} 0.412 & {\cellcolor[HTML]{087432}} \color[HTML]{F1F1F1} 0.334 & {\cellcolor[HTML]{248C46}} \color[HTML]{F1F1F1} 0.447 & {\cellcolor[HTML]{0B7734}} \color[HTML]{F1F1F1} 0.456 & {\cellcolor[HTML]{0D7836}} \color[HTML]{F1F1F1} 0.363 \\
ps \cite{3DS2A} & {\cellcolor[HTML]{2F984F}} \color[HTML]{F1F1F1} 0.391 & {\cellcolor[HTML]{157F3B}} \color[HTML]{F1F1F1} 0.401 & {\cellcolor[HTML]{0B7734}} \color[HTML]{F1F1F1} 0.330 & {\cellcolor[HTML]{228A44}} \color[HTML]{F1F1F1} 0.453 & {\cellcolor[HTML]{05712F}} \color[HTML]{F1F1F1} 0.468 & {\cellcolor[HTML]{05712F}} \color[HTML]{F1F1F1} 0.374 \\
run3 \cite{RACOON} & {\cellcolor[HTML]{006D2C}} \color[HTML]{F1F1F1} 0.481 & {\cellcolor[HTML]{1A843F}} \color[HTML]{F1F1F1} 0.392 & {\cellcolor[HTML]{0E7936}} \color[HTML]{F1F1F1} 0.326 & {\cellcolor[HTML]{006729}} \color[HTML]{F1F1F1} 0.530 & {\cellcolor[HTML]{0B7734}} \color[HTML]{F1F1F1} 0.455 & {\cellcolor[HTML]{0A7633}} \color[HTML]{F1F1F1} 0.367 \\
baseline-relevance-feedback \cite{CIRcluster} & {\cellcolor[HTML]{18823D}} \color[HTML]{F1F1F1} 0.436 & {\cellcolor[HTML]{067230}} \color[HTML]{F1F1F1} 0.427 & {\cellcolor[HTML]{157F3B}} \color[HTML]{F1F1F1} 0.319 & {\cellcolor[HTML]{0C7735}} \color[HTML]{F1F1F1} 0.494 & {\cellcolor[HTML]{006428}} \color[HTML]{F1F1F1} 0.490 & {\cellcolor[HTML]{0B7734}} \color[HTML]{F1F1F1} 0.366 \\
run4 \cite{RACOON} & {\cellcolor[HTML]{38A156}} \color[HTML]{F1F1F1} 0.374 & {\cellcolor[HTML]{1F8742}} \color[HTML]{F1F1F1} 0.383 & {\cellcolor[HTML]{16803C}} \color[HTML]{F1F1F1} 0.318 & {\cellcolor[HTML]{278F48}} \color[HTML]{F1F1F1} 0.441 & {\cellcolor[HTML]{0C7735}} \color[HTML]{F1F1F1} 0.454 & {\cellcolor[HTML]{0C7735}} \color[HTML]{F1F1F1} 0.365 \\
run2 \cite{RACOON} & {\cellcolor[HTML]{00692A}} \color[HTML]{F1F1F1} 0.487 & {\cellcolor[HTML]{1F8742}} \color[HTML]{F1F1F1} 0.383 & {\cellcolor[HTML]{16803C}} \color[HTML]{F1F1F1} 0.318 & {\cellcolor[HTML]{005F26}} \color[HTML]{F1F1F1} 0.543 & {\cellcolor[HTML]{0C7735}} \color[HTML]{F1F1F1} 0.454 & {\cellcolor[HTML]{0C7735}} \color[HTML]{F1F1F1} 0.365 \\
seupd2425-sard-minAnaMixed3-GPT \cite{SARD} & {\cellcolor[HTML]{359E53}} \color[HTML]{F1F1F1} 0.380 & {\cellcolor[HTML]{1E8741}} \color[HTML]{F1F1F1} 0.385 & {\cellcolor[HTML]{19833E}} \color[HTML]{F1F1F1} 0.314 & {\cellcolor[HTML]{268E47}} \color[HTML]{F1F1F1} 0.444 & {\cellcolor[HTML]{0D7836}} \color[HTML]{F1F1F1} 0.452 & {\cellcolor[HTML]{117B38}} \color[HTML]{F1F1F1} 0.359 \\
query-variants-qrel-boost-dbscan \cite{CIRcluster} & {\cellcolor[HTML]{208843}} \color[HTML]{F1F1F1} 0.422 & {\cellcolor[HTML]{157F3B}} \color[HTML]{F1F1F1} 0.400 & {\cellcolor[HTML]{208843}} \color[HTML]{F1F1F1} 0.307 & {\cellcolor[HTML]{17813D}} \color[HTML]{F1F1F1} 0.474 & {\cellcolor[HTML]{097532}} \color[HTML]{F1F1F1} 0.460 & {\cellcolor[HTML]{157F3B}} \color[HTML]{F1F1F1} 0.354 \\
query-variants-qrel-boost \cite{CIRcluster} & {\cellcolor[HTML]{208843}} \color[HTML]{F1F1F1} 0.422 & {\cellcolor[HTML]{157F3B}} \color[HTML]{F1F1F1} 0.400 & {\cellcolor[HTML]{208843}} \color[HTML]{F1F1F1} 0.307 & {\cellcolor[HTML]{17813D}} \color[HTML]{F1F1F1} 0.474 & {\cellcolor[HTML]{097532}} \color[HTML]{F1F1F1} 0.460 & {\cellcolor[HTML]{157F3B}} \color[HTML]{F1F1F1} 0.354 \\
seupd2425-sard-minAnaMixed \cite{SARD} & {\cellcolor[HTML]{3BA458}} \color[HTML]{F1F1F1} 0.367 & {\cellcolor[HTML]{248C46}} \color[HTML]{F1F1F1} 0.374 & {\cellcolor[HTML]{208843}} \color[HTML]{F1F1F1} 0.307 & {\cellcolor[HTML]{2D954D}} \color[HTML]{F1F1F1} 0.428 & {\cellcolor[HTML]{147E3A}} \color[HTML]{F1F1F1} 0.438 & {\cellcolor[HTML]{17813D}} \color[HTML]{F1F1F1} 0.351 \\
dsgt-bm25-expanded-rerank-submission \cite{DS@GT} & {\cellcolor[HTML]{3BA458}} \color[HTML]{F1F1F1} 0.367 & {\cellcolor[HTML]{248C46}} \color[HTML]{F1F1F1} 0.374 & {\cellcolor[HTML]{218944}} \color[HTML]{F1F1F1} 0.306 & {\cellcolor[HTML]{3AA357}} \color[HTML]{F1F1F1} 0.398 & {\cellcolor[HTML]{238B45}} \color[HTML]{F1F1F1} 0.410 & {\cellcolor[HTML]{278F48}} \color[HTML]{F1F1F1} 0.331 \\
JMFT\_lenght \cite{CIR} & {\cellcolor[HTML]{2A924A}} \color[HTML]{F1F1F1} 0.403 & {\cellcolor[HTML]{258D47}} \color[HTML]{F1F1F1} 0.373 & {\cellcolor[HTML]{278F48}} \color[HTML]{F1F1F1} 0.299 & {\cellcolor[HTML]{208843}} \color[HTML]{F1F1F1} 0.456 & {\cellcolor[HTML]{157F3B}} \color[HTML]{F1F1F1} 0.436 & {\cellcolor[HTML]{1C8540}} \color[HTML]{F1F1F1} 0.346 \\
JMFT\_Bert \cite{CIR} & {\cellcolor[HTML]{2A924A}} \color[HTML]{F1F1F1} 0.403 & {\cellcolor[HTML]{268E47}} \color[HTML]{F1F1F1} 0.370 & {\cellcolor[HTML]{278F48}} \color[HTML]{F1F1F1} 0.299 & {\cellcolor[HTML]{208843}} \color[HTML]{F1F1F1} 0.456 & {\cellcolor[HTML]{17813D}} \color[HTML]{F1F1F1} 0.433 & {\cellcolor[HTML]{1C8540}} \color[HTML]{F1F1F1} 0.346 \\
ci \cite{3DS2A} & {\cellcolor[HTML]{43AC5E}} \color[HTML]{F1F1F1} 0.351 & {\cellcolor[HTML]{2A924A}} \color[HTML]{F1F1F1} 0.362 & {\cellcolor[HTML]{278F48}} \color[HTML]{F1F1F1} 0.299 & {\cellcolor[HTML]{319A50}} \color[HTML]{F1F1F1} 0.418 & {\cellcolor[HTML]{17813D}} \color[HTML]{F1F1F1} 0.433 & {\cellcolor[HTML]{1C8540}} \color[HTML]{F1F1F1} 0.346 \\
JMFT\_Jaccard \cite{CIR} & {\cellcolor[HTML]{2A924A}} \color[HTML]{F1F1F1} 0.403 & {\cellcolor[HTML]{258D47}} \color[HTML]{F1F1F1} 0.373 & {\cellcolor[HTML]{289049}} \color[HTML]{F1F1F1} 0.297 & {\cellcolor[HTML]{208843}} \color[HTML]{F1F1F1} 0.456 & {\cellcolor[HTML]{157F3B}} \color[HTML]{F1F1F1} 0.436 & {\cellcolor[HTML]{1E8741}} \color[HTML]{F1F1F1} 0.343 \\
ci-prf-ps \cite{3DS2A} & {\cellcolor[HTML]{48AE60}} \color[HTML]{F1F1F1} 0.346 & {\cellcolor[HTML]{2B934B}} \color[HTML]{F1F1F1} 0.360 & {\cellcolor[HTML]{29914A}} \color[HTML]{F1F1F1} 0.296 & {\cellcolor[HTML]{339C52}} \color[HTML]{F1F1F1} 0.414 & {\cellcolor[HTML]{18823D}} \color[HTML]{F1F1F1} 0.431 & {\cellcolor[HTML]{1F8742}} \color[HTML]{F1F1F1} 0.342 \\
ps-prf \cite{3DS2A} & {\cellcolor[HTML]{4EB264}} \color[HTML]{F1F1F1} 0.337 & {\cellcolor[HTML]{309950}} \color[HTML]{F1F1F1} 0.350 & {\cellcolor[HTML]{2D954D}} \color[HTML]{F1F1F1} 0.291 & {\cellcolor[HTML]{37A055}} \color[HTML]{F1F1F1} 0.405 & {\cellcolor[HTML]{1C8540}} \color[HTML]{F1F1F1} 0.424 & {\cellcolor[HTML]{218944}} \color[HTML]{F1F1F1} 0.339 \\
dsgt-bm25-submission \cite{DS@GT} & {\cellcolor[HTML]{53B466}} \color[HTML]{F1F1F1} 0.331 & {\cellcolor[HTML]{359E53}} \color[HTML]{F1F1F1} 0.341 & {\cellcolor[HTML]{359E53}} \color[HTML]{F1F1F1} 0.280 & {\cellcolor[HTML]{3FA95C}} \color[HTML]{F1F1F1} 0.385 & {\cellcolor[HTML]{289049}} \color[HTML]{F1F1F1} 0.399 & {\cellcolor[HTML]{2F984F}} \color[HTML]{F1F1F1} 0.320 \\
molten-marsanne \cite{CIR} & {\cellcolor[HTML]{56B567}} \color[HTML]{F1F1F1} 0.326 & {\cellcolor[HTML]{3DA65A}} \color[HTML]{F1F1F1} 0.325 & {\cellcolor[HTML]{3FA85B}} \color[HTML]{F1F1F1} 0.268 & {\cellcolor[HTML]{3AA357}} \color[HTML]{F1F1F1} 0.398 & {\cellcolor[HTML]{278F48}} \color[HTML]{F1F1F1} 0.401 & {\cellcolor[HTML]{2F974E}} \color[HTML]{F1F1F1} 0.321 \\
datahunter-web-20250430-run1 \cite{DataHunter} & {\cellcolor[HTML]{60BA6C}} \color[HTML]{F1F1F1} 0.315 & {\cellcolor[HTML]{42AB5D}} \color[HTML]{F1F1F1} 0.316 & {\cellcolor[HTML]{4AAF61}} \color[HTML]{F1F1F1} 0.259 & {\cellcolor[HTML]{46AE60}} \color[HTML]{F1F1F1} 0.376 & {\cellcolor[HTML]{2F974E}} \color[HTML]{F1F1F1} 0.384 & {\cellcolor[HTML]{3AA357}} \color[HTML]{F1F1F1} 0.305 \\
searchill\_YYYY-MM [searchill] & {\cellcolor[HTML]{72C375}} \color[HTML]{000000} 0.293 & {\cellcolor[HTML]{4BB062}} \color[HTML]{F1F1F1} 0.304 & {\cellcolor[HTML]{4DB163}} \color[HTML]{F1F1F1} 0.256 & {\cellcolor[HTML]{50B264}} \color[HTML]{F1F1F1} 0.363 & {\cellcolor[HTML]{319A50}} \color[HTML]{F1F1F1} 0.379 & {\cellcolor[HTML]{38A156}} \color[HTML]{F1F1F1} 0.308 \\
seupd2425-sard-minAnaPhr300 \cite{SARD} & {\cellcolor[HTML]{63BC6E}} \color[HTML]{F1F1F1} 0.311 & {\cellcolor[HTML]{43AC5E}} \color[HTML]{F1F1F1} 0.314 & {\cellcolor[HTML]{4DB163}} \color[HTML]{F1F1F1} 0.256 & {\cellcolor[HTML]{60BA6C}} \color[HTML]{F1F1F1} 0.340 & {\cellcolor[HTML]{43AC5E}} \color[HTML]{F1F1F1} 0.342 & {\cellcolor[HTML]{5DB96B}} \color[HTML]{F1F1F1} 0.272 \\
query\_expansion\_time\_dependence \cite{CIR} & {\cellcolor[HTML]{78C679}} \color[HTML]{000000} 0.285 & {\cellcolor[HTML]{55B567}} \color[HTML]{F1F1F1} 0.293 & {\cellcolor[HTML]{50B264}} \color[HTML]{F1F1F1} 0.254 & {\cellcolor[HTML]{55B567}} \color[HTML]{F1F1F1} 0.355 & {\cellcolor[HTML]{37A055}} \color[HTML]{F1F1F1} 0.367 & {\cellcolor[HTML]{39A257}} \color[HTML]{F1F1F1} 0.307 \\
baseline-bm25 \cite{CIRcluster} & {\cellcolor[HTML]{7AC77B}} \color[HTML]{000000} 0.282 & {\cellcolor[HTML]{55B567}} \color[HTML]{F1F1F1} 0.293 & {\cellcolor[HTML]{50B264}} \color[HTML]{F1F1F1} 0.254 & {\cellcolor[HTML]{58B668}} \color[HTML]{F1F1F1} 0.352 & {\cellcolor[HTML]{37A055}} \color[HTML]{F1F1F1} 0.367 & {\cellcolor[HTML]{39A257}} \color[HTML]{F1F1F1} 0.307 \\
clef-webbusters-traditional [web-busters] & {\cellcolor[HTML]{7AC77B}} \color[HTML]{000000} 0.282 & {\cellcolor[HTML]{58B668}} \color[HTML]{F1F1F1} 0.290 & {\cellcolor[HTML]{55B567}} \color[HTML]{F1F1F1} 0.251 & {\cellcolor[HTML]{58B668}} \color[HTML]{F1F1F1} 0.352 & {\cellcolor[HTML]{37A055}} \color[HTML]{F1F1F1} 0.366 & {\cellcolor[HTML]{3AA357}} \color[HTML]{F1F1F1} 0.305 \\
seupd2425-sard-minAnaBool1000-GPT \cite{SARD} & {\cellcolor[HTML]{66BD6F}} \color[HTML]{F1F1F1} 0.307 & {\cellcolor[HTML]{48AE60}} \color[HTML]{F1F1F1} 0.308 & {\cellcolor[HTML]{55B567}} \color[HTML]{F1F1F1} 0.250 & {\cellcolor[HTML]{46AE60}} \color[HTML]{F1F1F1} 0.376 & {\cellcolor[HTML]{2F984F}} \color[HTML]{F1F1F1} 0.382 & {\cellcolor[HTML]{3DA65A}} \color[HTML]{F1F1F1} 0.301 \\
query-variants-relevance-feedback-dbscan \cite{CIRcluster} & {\cellcolor[HTML]{66BD6F}} \color[HTML]{F1F1F1} 0.307 & {\cellcolor[HTML]{4BB062}} \color[HTML]{F1F1F1} 0.305 & {\cellcolor[HTML]{5DB96B}} \color[HTML]{F1F1F1} 0.244 & {\cellcolor[HTML]{45AD5F}} \color[HTML]{F1F1F1} 0.377 & {\cellcolor[HTML]{309950}} \color[HTML]{F1F1F1} 0.381 & {\cellcolor[HTML]{3FA85B}} \color[HTML]{F1F1F1} 0.299 \\
fair\_schaer \cite{CIR} & {\cellcolor[HTML]{80CA80}} \color[HTML]{000000} 0.272 & {\cellcolor[HTML]{66BD6F}} \color[HTML]{F1F1F1} 0.272 & {\cellcolor[HTML]{73C476}} \color[HTML]{000000} 0.228 & {\cellcolor[HTML]{60BA6C}} \color[HTML]{F1F1F1} 0.340 & {\cellcolor[HTML]{42AB5D}} \color[HTML]{F1F1F1} 0.344 & {\cellcolor[HTML]{52B365}} \color[HTML]{F1F1F1} 0.282 \\
seupd-2425-basette \cite{BASETTE} & {\cellcolor[HTML]{7CC87C}} \color[HTML]{000000} 0.278 & {\cellcolor[HTML]{63BC6E}} \color[HTML]{F1F1F1} 0.276 & {\cellcolor[HTML]{76C578}} \color[HTML]{000000} 0.225 & {\cellcolor[HTML]{7FC97F}} \color[HTML]{000000} 0.297 & {\cellcolor[HTML]{68BE70}} \color[HTML]{000000} 0.295 & {\cellcolor[HTML]{84CC83}} \color[HTML]{000000} 0.238 \\
dsgt-bm25-expanded-submission \cite{DS@GT} & {\cellcolor[HTML]{8BCF89}} \color[HTML]{000000} 0.257 & {\cellcolor[HTML]{6EC173}} \color[HTML]{000000} 0.263 & {\cellcolor[HTML]{80CA80}} \color[HTML]{000000} 0.217 & {\cellcolor[HTML]{75C477}} \color[HTML]{000000} 0.313 & {\cellcolor[HTML]{52B365}} \color[HTML]{F1F1F1} 0.323 & {\cellcolor[HTML]{6ABF71}} \color[HTML]{000000} 0.262 \\
query-variants-relevance-feedback-kmeans \cite{CIRcluster} & {\cellcolor[HTML]{79C67A}} \color[HTML]{000000} 0.283 & {\cellcolor[HTML]{6DC072}} \color[HTML]{000000} 0.265 & {\cellcolor[HTML]{81CA81}} \color[HTML]{000000} 0.216 & {\cellcolor[HTML]{56B567}} \color[HTML]{F1F1F1} 0.354 & {\cellcolor[HTML]{42AB5D}} \color[HTML]{F1F1F1} 0.344 & {\cellcolor[HTML]{5BB86A}} \color[HTML]{F1F1F1} 0.274 \\
bm25\_boosted\_temporal\_categories \cite{CIR} & {\cellcolor[HTML]{A7DBA0}} \color[HTML]{000000} 0.216 & {\cellcolor[HTML]{88CE87}} \color[HTML]{000000} 0.229 & {\cellcolor[HTML]{90D18D}} \color[HTML]{000000} 0.203 & {\cellcolor[HTML]{81CA81}} \color[HTML]{000000} 0.294 & {\cellcolor[HTML]{5BB86A}} \color[HTML]{F1F1F1} 0.311 & {\cellcolor[HTML]{68BE70}} \color[HTML]{000000} 0.263 \\
seupd2425-rand-ICU-run \cite{RAND} & {\cellcolor[HTML]{B2E0AC}} \color[HTML]{000000} 0.196 & {\cellcolor[HTML]{9FD899}} \color[HTML]{000000} 0.199 & {\cellcolor[HTML]{BAE3B3}} \color[HTML]{000000} 0.164 & {\cellcolor[HTML]{98D594}} \color[HTML]{000000} 0.260 & {\cellcolor[HTML]{7CC87C}} \color[HTML]{000000} 0.267 & {\cellcolor[HTML]{9CD797}} \color[HTML]{000000} 0.215 \\
seupd2425-rand-frenchfilter-run \cite{RAND} & {\cellcolor[HTML]{BAE3B3}} \color[HTML]{000000} 0.185 & {\cellcolor[HTML]{A4DA9E}} \color[HTML]{000000} 0.190 & {\cellcolor[HTML]{C0E6B9}} \color[HTML]{000000} 0.158 & {\cellcolor[HTML]{9FD899}} \color[HTML]{000000} 0.248 & {\cellcolor[HTML]{83CB82}} \color[HTML]{000000} 0.257 & {\cellcolor[HTML]{A4DA9E}} \color[HTML]{000000} 0.208 \\
seupd2425-rand-DL-run \cite{RAND} & {\cellcolor[HTML]{BDE5B6}} \color[HTML]{000000} 0.179 & {\cellcolor[HTML]{A8DCA2}} \color[HTML]{000000} 0.185 & {\cellcolor[HTML]{C2E7BB}} \color[HTML]{000000} 0.155 & {\cellcolor[HTML]{AFDFA8}} \color[HTML]{000000} 0.220 & {\cellcolor[HTML]{95D391}} \color[HTML]{000000} 0.229 & {\cellcolor[HTML]{B5E1AE}} \color[HTML]{000000} 0.189 \\
seupd2425-rand-elision+synonyms0.5 \cite{RAND} & {\cellcolor[HTML]{C0E6B9}} \color[HTML]{000000} 0.176 & {\cellcolor[HTML]{ABDDA5}} \color[HTML]{000000} 0.180 & {\cellcolor[HTML]{C7E9C0}} \color[HTML]{000000} 0.150 & {\cellcolor[HTML]{A5DB9F}} \color[HTML]{000000} 0.238 & {\cellcolor[HTML]{8ACE88}} \color[HTML]{000000} 0.246 & {\cellcolor[HTML]{AADDA4}} \color[HTML]{000000} 0.201 \\
bm25-baseline+traditional [air5] & {\cellcolor[HTML]{C7E9C0}} \color[HTML]{000000} 0.163 & {\cellcolor[HTML]{B6E2AF}} \color[HTML]{000000} 0.163 & {\cellcolor[HTML]{CBEBC5}} \color[HTML]{000000} 0.144 & {\cellcolor[HTML]{C7E9C0}} \color[HTML]{000000} 0.179 & {\cellcolor[HTML]{B2E0AC}} \color[HTML]{000000} 0.183 & {\cellcolor[HTML]{CEECC8}} \color[HTML]{000000} 0.159 \\
luyuHF [air5] & {\cellcolor[HTML]{D1EDCB}} \color[HTML]{000000} 0.142 & {\cellcolor[HTML]{F7FCF5}} \color[HTML]{000000} 0.014 & {\cellcolor[HTML]{D2EDCC}} \color[HTML]{000000} 0.136 & {\cellcolor[HTML]{CDECC7}} \color[HTML]{000000} 0.163 & {\cellcolor[HTML]{F7FCF5}} \color[HTML]{000000} 0.014 & {\cellcolor[HTML]{DEF2D9}} \color[HTML]{000000} 0.136 \\
dense-baseline+multilingual [air5] & {\cellcolor[HTML]{F7FCF5}} \color[HTML]{000000} 0.034 & {\cellcolor[HTML]{F1FAEE}} \color[HTML]{000000} 0.033 & {\cellcolor[HTML]{F7FCF5}} \color[HTML]{000000} 0.073 & {\cellcolor[HTML]{F7FCF5}} \color[HTML]{000000} 0.040 & {\cellcolor[HTML]{F0F9ED}} \color[HTML]{000000} 0.040 & {\cellcolor[HTML]{F7FCF5}} \color[HTML]{000000} 0.084 \\
\bottomrule
\end{tabular}
}
\end{table}

\subsection{Changes in Effectiveness}
Especially interesting to this lab is to investigate how the retrieval effectiveness changes over time and which approaches are more resilient against an evolving search setting. Like in previous iterations of this lab, the Relative Improvement (RI) in effectiveness is measured~\cite{longevaloverview2023,longevalCLEF2024overview}. Additionally, we employ the Delta Relative Improvement (DRI) and Effect Ratio (ER) to further describe the changes in the effectiveness with a focus on the system effect~\cite{DBLP:conf/sigir/Breuer0FMSSS20,keller2024evaluation,keller2024leveraging}. Both measures first relate the measured scores to a pivot system, BM25~\cite{robertsonOkapiTREC31994} in our case, and then compare the emerging deltas across time. 
Since the experimental system of interest and the pivot system are exposed to the same changes, the changes introduced by the evolving search setting are dampened, and the results emphasize the effect of the experimental system. 
This is especially of interest when the experimental system is updated, retrained, or directly interacts with the previous snapshots. In these cases, a robust system effect is desirable. 
The change and persistence-based measures are complemented with the Average Retrieval Performance (ARP) across all topics at the evolved snapshot based on the instantiation measure nDCG@10 and the Mean Average Retrieval Performance (MARP) that calculates the mean between both compared snapshots.

For the Web task, short-term and long-term changes are compared and the results are displayed in the Tables~\ref{tab:web-results-change-long} and \ref{tab:web-results-change-short}. The reference system BM25 did not rank any results for up to 73 topics. To calculate the per-topic differences in the experimental systems, these topics are removed. Therefore, the ARP may vary slightly between the Tables~\ref{tab:web-results}, \ref{tab:web-results-change-long}, and \ref{tab:web-results-change-short}.
The RI and DRI measure how the effectiveness changes relative to the first snapshot 2023-03. Due to their definition, values below zero indicate improving effectiveness. Almost half of the systems show an improved effectiveness over the short timespan. In contrast, only the three approaches \texttt{ci-prf-ps}, \texttt{longeval\_bm25\_boosted\_temporal\_categories}, and \texttt{luyuHF} indicate an improving system effect. This disagreement is also reflected in the top three approaches with the lowest change. RI rank the approaches \texttt{seupd2425-rand-ICU-run}, and \texttt{seupd2425-rand-{\allowbreak}elision+synonyms0.5} followed by \texttt{dsgt-bm25-rerank-submission} and \texttt{seupd2425-sard-minAnaBool1000-GPT} with an RI of 0.003.
The systems \texttt{ci-prf-ps}, \texttt{ps-prf}, and \texttt{searchill\_YYYY-MM} indicate the lowest system effect change according to DRI. These systems also show a low RI. Since the effectiveness is generally not changing much, all differences are small.
Regarding the long-term changes, only the \texttt{dense-baseline+multilingual} improves in effectiveness measured by RI. The approaches \texttt{luyuHF}, \texttt{longeval\_bm25\_{\allowbreak}boosted\_{\allowbreak}temporal\_categories} and \texttt{query\_expansion\_time\_dependence} show the lowest changes. Regarding the system effects measured by DRI, almost all approaches impair again. \texttt{query\_expansion\_time\_dependence}, \texttt{mein-cleveres-system}, and \texttt{clef-webbusters-traditional} and seem to be most robust.

\begin{table}[h]
\caption{Evaluation results for the SciRetrieval task. The results are sorted by nDCG@10 for the 2025-01 snapshot.}
\label{tab:sci-results}
\centering
\resizebox{\textwidth}{!}{%
\begin{tabular}{lcc@{\hskip 5pt}cc}
\toprule
 & \multicolumn{2}{c}{nDCG@10} & \multicolumn{2}{c}{nDCG@1000} \\
 Approach  & 2024-11 & 2025-01 & 2024-11 & 2025-01 \\
 \midrule
BM25\_k1\_1p0\_b\_0p7\_stop\_stem\_fullText [academy-retrievals] & {\cellcolor[HTML]{3FA85B}} \color[HTML]{F1F1F1} 0.102 & {\cellcolor[HTML]{00441B}} \color[HTML]{F1F1F1} 0.292 & {\cellcolor[HTML]{84CC83}} \color[HTML]{000000} 0.156 & {\cellcolor[HTML]{00441B}} \color[HTML]{F1F1F1} 0.393 \\
BM25\_k1\_0p2\_b\_0p8\_stop\_stem\_fullText [academy-retrievals] & {\cellcolor[HTML]{3FA85B}} \color[HTML]{F1F1F1} 0.102 & {\cellcolor[HTML]{00441B}} \color[HTML]{F1F1F1} 0.291 & {\cellcolor[HTML]{87CD86}} \color[HTML]{000000} 0.154 & {\cellcolor[HTML]{00451C}} \color[HTML]{F1F1F1} 0.391 \\
BM25\_k1\_0p95\_b\_0p75\_stop\_stem\_fullText [academy-retrievals] & {\cellcolor[HTML]{3EA75A}} \color[HTML]{F1F1F1} 0.103 & {\cellcolor[HTML]{00451C}} \color[HTML]{F1F1F1} 0.290 & {\cellcolor[HTML]{84CC83}} \color[HTML]{000000} 0.156 & {\cellcolor[HTML]{00451C}} \color[HTML]{F1F1F1} 0.391 \\
rm3-on-qrel-boost \cite{OWS} & {\cellcolor[HTML]{005A24}} \color[HTML]{F1F1F1} 0.150 & {\cellcolor[HTML]{03702E}} \color[HTML]{F1F1F1} 0.252 & {\cellcolor[HTML]{005020}} \color[HTML]{F1F1F1} 0.331 & {\cellcolor[HTML]{004A1E}} \color[HTML]{F1F1F1} 0.384 \\
qrel-boost-core \cite{OWS} & {\cellcolor[HTML]{A0D99B}} \color[HTML]{000000} 0.061 & {\cellcolor[HTML]{17813D}} \color[HTML]{F1F1F1} 0.231 & {\cellcolor[HTML]{16803C}} \color[HTML]{F1F1F1} 0.274 & {\cellcolor[HTML]{00441B}} \color[HTML]{F1F1F1} 0.393 \\
query-intent-fusion \cite{OWS} & {\cellcolor[HTML]{A0D99B}} \color[HTML]{000000} 0.061 & {\cellcolor[HTML]{17813D}} \color[HTML]{F1F1F1} 0.231 & {\cellcolor[HTML]{16803C}} \color[HTML]{F1F1F1} 0.274 & {\cellcolor[HTML]{00441B}} \color[HTML]{F1F1F1} 0.393 \\
ows-cluster-boosting \cite{OWS} & {\cellcolor[HTML]{00441B}} \color[HTML]{F1F1F1} 0.161 & {\cellcolor[HTML]{228A44}} \color[HTML]{F1F1F1} 0.220 & {\cellcolor[HTML]{00441B}} \color[HTML]{F1F1F1} 0.344 & {\cellcolor[HTML]{00682A}} \color[HTML]{F1F1F1} 0.350 \\
ows-bm25 \cite{OWS} & {\cellcolor[HTML]{00441B}} \color[HTML]{F1F1F1} 0.161 & {\cellcolor[HTML]{258D47}} \color[HTML]{F1F1F1} 0.216 & {\cellcolor[HTML]{00441B}} \color[HTML]{F1F1F1} 0.344 & {\cellcolor[HTML]{00692A}} \color[HTML]{F1F1F1} 0.348 \\
bm25+reranker [tf-idk] & {\cellcolor[HTML]{8ED08B}} \color[HTML]{000000} 0.069 & {\cellcolor[HTML]{40AA5D}} \color[HTML]{F1F1F1} 0.183 & {\cellcolor[HTML]{A8DCA2}} \color[HTML]{000000} 0.121 & {\cellcolor[HTML]{3DA65A}} \color[HTML]{F1F1F1} 0.253 \\
bm25+reranker+weighted [tf-idk] & {\cellcolor[HTML]{8ACE88}} \color[HTML]{000000} 0.071 & {\cellcolor[HTML]{4AAF61}} \color[HTML]{F1F1F1} 0.176 & {\cellcolor[HTML]{A4DA9E}} \color[HTML]{000000} 0.126 & {\cellcolor[HTML]{3CA559}} \color[HTML]{F1F1F1} 0.254 \\
BM25\_ColBERTv2\_ReRanker\_v1 [academy-retrievals] & {\cellcolor[HTML]{A2D99C}} \color[HTML]{000000} 0.060 & {\cellcolor[HTML]{4BB062}} \color[HTML]{F1F1F1} 0.175 & {\cellcolor[HTML]{A4DA9E}} \color[HTML]{000000} 0.125 & {\cellcolor[HTML]{2A924A}} \color[HTML]{F1F1F1} 0.283 \\
bm25 [tf-idk] & {\cellcolor[HTML]{0A7633}} \color[HTML]{F1F1F1} 0.135 & {\cellcolor[HTML]{50B264}} \color[HTML]{F1F1F1} 0.172 & {\cellcolor[HTML]{2E964D}} \color[HTML]{F1F1F1} 0.242 & {\cellcolor[HTML]{42AB5D}} \color[HTML]{F1F1F1} 0.245 \\
A3-TFIDF-TestAbs [sambs] & {\cellcolor[HTML]{90D18D}} \color[HTML]{000000} 0.068 & {\cellcolor[HTML]{8DD08A}} \color[HTML]{000000} 0.126 & {\cellcolor[HTML]{AADDA4}} \color[HTML]{000000} 0.119 & {\cellcolor[HTML]{6DC072}} \color[HTML]{000000} 0.204 \\
A4-TFIDF-TestFull [sambs] & {\cellcolor[HTML]{95D391}} \color[HTML]{000000} 0.066 & {\cellcolor[HTML]{91D28E}} \color[HTML]{000000} 0.123 & {\cellcolor[HTML]{ABDDA5}} \color[HTML]{000000} 0.117 & {\cellcolor[HTML]{70C274}} \color[HTML]{000000} 0.201 \\
B4-SBERT-V2-MPNet-TestFull [sambs] & {\cellcolor[HTML]{DFF3DA}} \color[HTML]{000000} 0.024 & {\cellcolor[HTML]{D0EDCA}} \color[HTML]{000000} 0.062 & {\cellcolor[HTML]{E5F5E0}} \color[HTML]{000000} 0.043 & {\cellcolor[HTML]{C8E9C1}} \color[HTML]{000000} 0.097 \\
B3-SBERT-V2-MPNet-TestAbs [sambs] & {\cellcolor[HTML]{E1F3DC}} \color[HTML]{000000} 0.023 & {\cellcolor[HTML]{D1EDCB}} \color[HTML]{000000} 0.061 & {\cellcolor[HTML]{E6F5E1}} \color[HTML]{000000} 0.041 & {\cellcolor[HTML]{CAEAC3}} \color[HTML]{000000} 0.095 \\
fusion-with-core \cite{OWS} & {\cellcolor[HTML]{A0D99B}} \color[HTML]{000000} 0.061 & {\cellcolor[HTML]{D3EECD}} \color[HTML]{000000} 0.059 & {\cellcolor[HTML]{16803C}} \color[HTML]{F1F1F1} 0.274 & {\cellcolor[HTML]{369F54}} \color[HTML]{F1F1F1} 0.263 \\
SciBERT\_LongEVAL [long-eval-sci-group-5] & {\cellcolor[HTML]{DAF0D4}} \color[HTML]{000000} 0.028 & {\cellcolor[HTML]{D3EECD}} \color[HTML]{000000} 0.059 & {\cellcolor[HTML]{DBF1D5}} \color[HTML]{000000} 0.059 & {\cellcolor[HTML]{B8E3B2}} \color[HTML]{000000} 0.117 \\
B3-SBERT-V1-TestAbs [sambs] & {\cellcolor[HTML]{DFF3DA}} \color[HTML]{000000} 0.024 & {\cellcolor[HTML]{D6EFD0}} \color[HTML]{000000} 0.055 & {\cellcolor[HTML]{E3F4DE}} \color[HTML]{000000} 0.046 & {\cellcolor[HTML]{CBEBC5}} \color[HTML]{000000} 0.092 \\
clef25-agh-cracow \cite{EAIiIB} & {\cellcolor[HTML]{F7FCF5}} \color[HTML]{000000} 0.000 & {\cellcolor[HTML]{E2F4DD}} \color[HTML]{000000} 0.041 & {\cellcolor[HTML]{F7FCF5}} \color[HTML]{000000} 0.000 & {\cellcolor[HTML]{E7F6E2}} \color[HTML]{000000} 0.045 \\
monot5-in-core \cite{OWS} & {\cellcolor[HTML]{DBF1D6}} \color[HTML]{000000} 0.027 & {\cellcolor[HTML]{E8F6E3}} \color[HTML]{000000} 0.031 & {\cellcolor[HTML]{37A055}} \color[HTML]{F1F1F1} 0.229 & {\cellcolor[HTML]{5DB96B}} \color[HTML]{F1F1F1} 0.218 \\
C3-CE-L6-TestAbs [sambs] & {\cellcolor[HTML]{F7FCF5}} \color[HTML]{000000} 0.000 & {\cellcolor[HTML]{F7FCF5}} \color[HTML]{000000} 0.000 & {\cellcolor[HTML]{F7FCF5}} \color[HTML]{000000} 0.000 & {\cellcolor[HTML]{F7FCF5}} \color[HTML]{000000} 0.000 \\
C3-CE-L12-TestAbs [sambs] & {\cellcolor[HTML]{F7FCF5}} \color[HTML]{000000} 0.000 & {\cellcolor[HTML]{F7FCF5}} \color[HTML]{000000} 0.000 & {\cellcolor[HTML]{F7FCF5}} \color[HTML]{000000} 0.000 & {\cellcolor[HTML]{F7FCF5}} \color[HTML]{000000} 0.000 \\
A7-BM25-TestAbs [sambs] & {\cellcolor[HTML]{F7FCF5}} \color[HTML]{000000} 0.000 & {\cellcolor[HTML]{F7FCF5}} \color[HTML]{000000} 0.000 & {\cellcolor[HTML]{F7FCF5}} \color[HTML]{000000} 0.000 & {\cellcolor[HTML]{F7FCF5}} \color[HTML]{000000} 0.000 \\
A8-BM25-TestFull [sambs] & {\cellcolor[HTML]{F7FCF5}} \color[HTML]{000000} 0.000 & {\cellcolor[HTML]{F7FCF5}} \color[HTML]{000000} 0.000 & {\cellcolor[HTML]{F7FCF5}} \color[HTML]{000000} 0.000 & {\cellcolor[HTML]{F7FCF5}} \color[HTML]{000000} 0.000 \\
B4-SBERT-V1-TestFull [sambs] & {\cellcolor[HTML]{F7FCF5}} \color[HTML]{000000} 0.000 & {\cellcolor[HTML]{F7FCF5}} \color[HTML]{000000} 0.000 & {\cellcolor[HTML]{F7FCF5}} \color[HTML]{000000} 0.000 & {\cellcolor[HTML]{F7FCF5}} \color[HTML]{000000} 0.000 \\
C4-CE-L6-TestFull [sambs] & {\cellcolor[HTML]{F7FCF5}} \color[HTML]{000000} 0.000 & {\cellcolor[HTML]{F7FCF5}} \color[HTML]{000000} 0.000 & {\cellcolor[HTML]{F7FCF5}} \color[HTML]{000000} 0.000 & {\cellcolor[HTML]{F7FCF5}} \color[HTML]{000000} 0.000 \\
C4-CE-L12-TestFull [sambs] & {\cellcolor[HTML]{F7FCF5}} \color[HTML]{000000} 0.000 & {\cellcolor[HTML]{F7FCF5}} \color[HTML]{000000} 0.000 & {\cellcolor[HTML]{F7FCF5}} \color[HTML]{000000} 0.000 & {\cellcolor[HTML]{F7FCF5}} \color[HTML]{000000} 0.000 \\
\bottomrule
\end{tabular}
}
\end{table}

\begin{table}[h]							
\caption{Standardized results for the SciRetrieval task. The results are sorted by nDCG@10 for the 2025-01 snapshot, as in Table~\ref{tab:sci-results}.}									
\label{tab:sci-results-std}			
\centering
\resizebox{0.9\textwidth}{!}{%
\begin{tabular}{lcc@{\hskip 5pt}cc}									
\toprule									
	&	\multicolumn{2}{c}{Emp(nDCG@10)}	&	\multicolumn{2}{c}{Emp(nDCG@1000)} \\					
Approach	&	2024-11	&	2025-01	&	2024-11	&	2025-01 \\	
\midrule									
BM25\_k1\_1p0\_b\_0p7\_stop\_stem\_fullText [academy-retrievals]	&	{\cellcolor[HTML]{2A924A}} \color[HTML]{F1F1F1} 0.818	&	{\cellcolor[HTML]{00471C}} \color[HTML]{F1F1F1} 0.836	&	{\cellcolor[HTML]{62BB6D}} \color[HTML]{F1F1F1} 0.654	&	{\cellcolor[HTML]{004E1F}} \color[HTML]{F1F1F1} 0.761 \\	
BM25\_k1\_0p2\_b\_0p8\_stop\_stem\_fullText [academy-retrievals]	&	{\cellcolor[HTML]{2F984F}} \color[HTML]{F1F1F1} 0.814	&	{\cellcolor[HTML]{005020}} \color[HTML]{F1F1F1} 0.829	&	{\cellcolor[HTML]{70C274}} \color[HTML]{000000} 0.637	&	{\cellcolor[HTML]{005C25}} \color[HTML]{F1F1F1} 0.741 \\	
BM25\_k1\_0p95\_b\_0p75\_stop\_stem\_fullText [academy-retrievals]	&	{\cellcolor[HTML]{278F48}} \color[HTML]{F1F1F1} 0.820	&	{\cellcolor[HTML]{00441B}} \color[HTML]{F1F1F1} 0.838	&	{\cellcolor[HTML]{62BB6D}} \color[HTML]{F1F1F1} 0.653	&	{\cellcolor[HTML]{005321}} \color[HTML]{F1F1F1} 0.755 \\	
rm3-on-qrel-boost \cite{OWS}	&	{\cellcolor[HTML]{05712F}} \color[HTML]{F1F1F1} 0.841	&	{\cellcolor[HTML]{03702E}} \color[HTML]{F1F1F1} 0.803	&	{\cellcolor[HTML]{006729}} \color[HTML]{F1F1F1} 0.813	&	{\cellcolor[HTML]{00441B}} \color[HTML]{F1F1F1} 0.776 \\	
qrel-boost-core \cite{OWS}	&	{\cellcolor[HTML]{90D18D}} \color[HTML]{000000} 0.769	&	{\cellcolor[HTML]{38A156}} \color[HTML]{F1F1F1} 0.751	&	{\cellcolor[HTML]{006227}} \color[HTML]{F1F1F1} 0.819	&	{\cellcolor[HTML]{004A1E}} \color[HTML]{F1F1F1} 0.767 \\	
query-intent-fusion \cite{OWS}	&	{\cellcolor[HTML]{90D18D}} \color[HTML]{000000} 0.769	&	{\cellcolor[HTML]{38A156}} \color[HTML]{F1F1F1} 0.751	&	{\cellcolor[HTML]{006227}} \color[HTML]{F1F1F1} 0.819	&	{\cellcolor[HTML]{004A1E}} \color[HTML]{F1F1F1} 0.767 \\	
ows-cluster-boosting \cite{OWS}	&	{\cellcolor[HTML]{00441B}} \color[HTML]{F1F1F1} 0.864	&	{\cellcolor[HTML]{006B2B}} \color[HTML]{F1F1F1} 0.807	&	{\cellcolor[HTML]{00441B}} \color[HTML]{F1F1F1} 0.863	&	{\cellcolor[HTML]{00441B}} \color[HTML]{F1F1F1} 0.777 \\	
ows-bm25 \cite{OWS}	&	{\cellcolor[HTML]{00441B}} \color[HTML]{F1F1F1} 0.864	&	{\cellcolor[HTML]{016E2D}} \color[HTML]{F1F1F1} 0.804	&	{\cellcolor[HTML]{00441B}} \color[HTML]{F1F1F1} 0.863	&	{\cellcolor[HTML]{00471C}} \color[HTML]{F1F1F1} 0.772 \\	
bm25+reranker [tf-idk]	&	{\cellcolor[HTML]{4BB062}} \color[HTML]{F1F1F1} 0.798	&	{\cellcolor[HTML]{3FA95C}} \color[HTML]{F1F1F1} 0.744	&	{\cellcolor[HTML]{87CD86}} \color[HTML]{000000} 0.609	&	{\cellcolor[HTML]{3EA75A}} \color[HTML]{F1F1F1} 0.611 \\	
bm25+reranker+weighted [tf-idk]	&	{\cellcolor[HTML]{58B668}} \color[HTML]{F1F1F1} 0.793	&	{\cellcolor[HTML]{3DA65A}} \color[HTML]{F1F1F1} 0.747	&	{\cellcolor[HTML]{8ACE88}} \color[HTML]{000000} 0.605	&	{\cellcolor[HTML]{3CA559}} \color[HTML]{F1F1F1} 0.613 \\	
BM25\_ColBERTv2\_ReRanker\_v1 [academy-retrievals]	&	{\cellcolor[HTML]{80CA80}} \color[HTML]{000000} 0.776	&	{\cellcolor[HTML]{46AE60}} \color[HTML]{F1F1F1} 0.739	&	{\cellcolor[HTML]{87CD86}} \color[HTML]{000000} 0.608	&	{\cellcolor[HTML]{2B934B}} \color[HTML]{F1F1F1} 0.646 \\	
bm25 [tf-idk]	&	{\cellcolor[HTML]{1D8640}} \color[HTML]{F1F1F1} 0.826	&	{\cellcolor[HTML]{42AB5D}} \color[HTML]{F1F1F1} 0.742	&	{\cellcolor[HTML]{3EA75A}} \color[HTML]{F1F1F1} 0.696	&	{\cellcolor[HTML]{45AD5F}} \color[HTML]{F1F1F1} 0.599 \\	
A3-TFIDF-TestAbs [sambs]	&	{\cellcolor[HTML]{4AAF61}} \color[HTML]{F1F1F1} 0.798	&	{\cellcolor[HTML]{78C679}} \color[HTML]{000000} 0.708	&	{\cellcolor[HTML]{86CC85}} \color[HTML]{000000} 0.610	&	{\cellcolor[HTML]{75C477}} \color[HTML]{000000} 0.545 \\	
A4-TFIDF-TestFull [sambs]	&	{\cellcolor[HTML]{4BB062}} \color[HTML]{F1F1F1} 0.798	&	{\cellcolor[HTML]{7DC87E}} \color[HTML]{000000} 0.704	&	{\cellcolor[HTML]{87CD86}} \color[HTML]{000000} 0.609	&	{\cellcolor[HTML]{79C67A}} \color[HTML]{000000} 0.540 \\	
B4-SBERT-V2-MPNet-TestFull [sambs]	&	{\cellcolor[HTML]{CCEBC6}} \color[HTML]{000000} 0.736	&	{\cellcolor[HTML]{CAEAC3}} \color[HTML]{000000} 0.644	&	{\cellcolor[HTML]{E4F5DF}} \color[HTML]{000000} 0.464	&	{\cellcolor[HTML]{C8E9C1}} \color[HTML]{000000} 0.431 \\	
B3-SBERT-V2-MPNet-TestAbs [sambs]	&	{\cellcolor[HTML]{CAEAC3}} \color[HTML]{000000} 0.738	&	{\cellcolor[HTML]{CCEBC6}} \color[HTML]{000000} 0.641	&	{\cellcolor[HTML]{E6F5E1}} \color[HTML]{000000} 0.458	&	{\cellcolor[HTML]{CAEAC3}} \color[HTML]{000000} 0.426 \\	
fusion-with-core \cite{OWS}	&	{\cellcolor[HTML]{90D18D}} \color[HTML]{000000} 0.769	&	{\cellcolor[HTML]{CEECC8}} \color[HTML]{000000} 0.639	&	{\cellcolor[HTML]{006227}} \color[HTML]{F1F1F1} 0.819	&	{\cellcolor[HTML]{208843}} \color[HTML]{F1F1F1} 0.666 \\	
SciBERT\_LongEVAL [long-eval-sci-group-5]	&	{\cellcolor[HTML]{C2E7BB}} \color[HTML]{000000} 0.742	&	{\cellcolor[HTML]{CDECC7}} \color[HTML]{000000} 0.640	&	{\cellcolor[HTML]{CFECC9}} \color[HTML]{000000} 0.502	&	{\cellcolor[HTML]{B5E1AE}} \color[HTML]{000000} 0.460 \\	
B3-SBERT-V1-TestAbs [sambs]	&	{\cellcolor[HTML]{D2EDCC}} \color[HTML]{000000} 0.732	&	{\cellcolor[HTML]{D7EFD1}} \color[HTML]{000000} 0.630	&	{\cellcolor[HTML]{E2F4DD}} \color[HTML]{000000} 0.466	&	{\cellcolor[HTML]{CCEBC6}} \color[HTML]{000000} 0.421 \\	
clef25-agh-cracow \cite{EAIiIB}	&	{\cellcolor[HTML]{F7FCF5}} \color[HTML]{000000} 0.698	&	{\cellcolor[HTML]{DBF1D5}} \color[HTML]{000000} 0.627	&	{\cellcolor[HTML]{F7FCF5}} \color[HTML]{000000} 0.403	&	{\cellcolor[HTML]{E6F5E1}} \color[HTML]{000000} 0.372 \\	
monot5-in-core \cite{OWS}	&	{\cellcolor[HTML]{D9F0D3}} \color[HTML]{000000} 0.727	&	{\cellcolor[HTML]{EBF7E7}} \color[HTML]{000000} 0.604	&	{\cellcolor[HTML]{5EB96B}} \color[HTML]{F1F1F1} 0.656	&	{\cellcolor[HTML]{81CA81}} \color[HTML]{000000} 0.529 \\	
C3-CE-L6-TestAbs [sambs]	&	{\cellcolor[HTML]{F7FCF5}} \color[HTML]{000000} 0.698	&	{\cellcolor[HTML]{F7FCF5}} \color[HTML]{000000} 0.583	&	{\cellcolor[HTML]{F7FCF5}} \color[HTML]{000000} 0.403	&	{\cellcolor[HTML]{F7FCF5}} \color[HTML]{000000} 0.316 \\	
C3-CE-L12-TestAbs [sambs]	&	{\cellcolor[HTML]{F7FCF5}} \color[HTML]{000000} 0.698	&	{\cellcolor[HTML]{F7FCF5}} \color[HTML]{000000} 0.583	&	{\cellcolor[HTML]{F7FCF5}} \color[HTML]{000000} 0.403	&	{\cellcolor[HTML]{F7FCF5}} \color[HTML]{000000} 0.316 \\	
A7-BM25-TestAbs [sambs]	&	{\cellcolor[HTML]{F7FCF5}} \color[HTML]{000000} 0.000	&	{\cellcolor[HTML]{F7FCF5}} \color[HTML]{000000} 0.000	&	{\cellcolor[HTML]{F7FCF5}} \color[HTML]{000000} 0.000	&	{\cellcolor[HTML]{F7FCF5}} \color[HTML]{000000} 0.000 \\	
A8-BM25-TestFull [sambs]	&	{\cellcolor[HTML]{F7FCF5}} \color[HTML]{000000} 0.000	&	{\cellcolor[HTML]{F7FCF5}} \color[HTML]{000000} 0.000	&	{\cellcolor[HTML]{F7FCF5}} \color[HTML]{000000} 0.000	&	{\cellcolor[HTML]{F7FCF5}} \color[HTML]{000000} 0.000 \\	
B4-SBERT-V1-TestFull [sambs]	&	{\cellcolor[HTML]{F7FCF5}} \color[HTML]{000000} 0.000	&	{\cellcolor[HTML]{F7FCF5}} \color[HTML]{000000} 0.000	&	{\cellcolor[HTML]{F7FCF5}} \color[HTML]{000000} 0.000	&	{\cellcolor[HTML]{F7FCF5}} \color[HTML]{000000} 0.000 \\	
C4-CE-L6-TestFull [sambs]	&	{\cellcolor[HTML]{F7FCF5}} \color[HTML]{000000} 0.000	&	{\cellcolor[HTML]{F7FCF5}} \color[HTML]{000000} 0.000	&	{\cellcolor[HTML]{F7FCF5}} \color[HTML]{000000} 0.000	&	{\cellcolor[HTML]{F7FCF5}} \color[HTML]{000000} 0.000 \\	
C4-CE-L12-TestFull [sambs]	&	{\cellcolor[HTML]{F7FCF5}} \color[HTML]{000000} 0.000	&	{\cellcolor[HTML]{F7FCF5}} \color[HTML]{000000} 0.000	&	{\cellcolor[HTML]{F7FCF5}} \color[HTML]{000000} 0.000	&	{\cellcolor[HTML]{F7FCF5}} \color[HTML]{000000} 0.000 \\	
\bottomrule									
\end{tabular}									
}									
\end{table}

\begin{table}[!htb]
\caption{Change results for the SciRetrieval task in comparison to the snapshot 2024-11. The results are sorted by RI. Darker colors indicate less change or higher effectiveness.}
\label{tab:sci-results-change}
\resizebox{\textwidth}{!}{%
\begin{tabular}{lc@{\hskip 5pt}c@{\hskip 5pt}c@{\hskip 5pt}c@{\hskip 5pt}c}
\toprule
Approach & ARP & AARP & RI & DRI & ER \\
\midrule
fusion-with-core \cite{OWS} & {\cellcolor[HTML]{D0E1F2}} \color[HTML]{000000} 0.059 & {\cellcolor[HTML]{B7D4EA}} \color[HTML]{000000} 0.060 & {\cellcolor[HTML]{9EB0FF}} \color{black} 0.029 & {\cellcolor[HTML]{9EB0FF}} \color{black} 0.108 & {\cellcolor[HTML]{111A20}} \color{white} 1.523 \\
monot5-in-core \cite{OWS} & {\cellcolor[HTML]{E2EDF8}} \color[HTML]{000000} 0.031 & {\cellcolor[HTML]{DBE9F6}} \color[HTML]{000000} 0.029 & {\cellcolor[HTML]{180C0A}} \color{white} -0.120 & {\cellcolor[HTML]{132D3A}} \color{white} 0.025 & {\cellcolor[HTML]{111317}} \color{white} 1.312 \\
ows-bm25 \cite{OWS} & {\cellcolor[HTML]{2171B5}} \color[HTML]{F1F1F1} 0.218 & {\cellcolor[HTML]{083C7D}} \color[HTML]{F1F1F1} 0.191 & {\cellcolor[HTML]{180C0A}} \color{white} -0.339 & {\cellcolor[HTML]{210B03}} \color{white} -0.058 & {\cellcolor[HTML]{112028}} \color{white} 1.673 \\
ows-cluster-boosting \cite{OWS} & {\cellcolor[HTML]{1F6EB3}} \color[HTML]{F1F1F1} 0.222 & {\cellcolor[HTML]{083A7A}} \color[HTML]{F1F1F1} 0.193 & {\cellcolor[HTML]{180C0A}} \color{white} -0.363 & {\cellcolor[HTML]{240C02}} \color{white} -0.080 & {\cellcolor[HTML]{112630}} \color{white} 1.816 \\
rm3-on-qrel-boost \cite{OWS} & {\cellcolor[HTML]{0A539E}} \color[HTML]{F1F1F1} 0.253 & {\cellcolor[HTML]{08306B}} \color[HTML]{F1F1F1} 0.201 & {\cellcolor[HTML]{180C0A}} \color{white} -0.680 & {\cellcolor[HTML]{591C07}} \color{white} -0.351 & {\cellcolor[HTML]{9EB0FF}} \color{black} 5.257 \\
A3-TFIDF-TestAbs [sambs] & {\cellcolor[HTML]{77B5D9}} \color[HTML]{000000} 0.137 & {\cellcolor[HTML]{69ADD5}} \color[HTML]{F1F1F1} 0.102 & {\cellcolor[HTML]{180C0A}} \color{white} -1.028 & {\cellcolor[HTML]{4D1602}} \color{white} -0.294 & {\cellcolor[HTML]{250C01}} \color{white} 0.585 \\
A4-TFIDF-TestFull [sambs] & {\cellcolor[HTML]{7CB7DA}} \color[HTML]{000000} 0.134 & {\cellcolor[HTML]{6CAED6}} \color[HTML]{F1F1F1} 0.100 & {\cellcolor[HTML]{180C0A}} \color{white} -1.028 & {\cellcolor[HTML]{4B1602}} \color{white} -0.287 & {\cellcolor[HTML]{240C02}} \color{white} 0.619 \\
SciBERT\_LongEVAL [long-eval-sci-group-5] & {\cellcolor[HTML]{D0E1F2}} \color[HTML]{000000} 0.059 & {\cellcolor[HTML]{CDE0F1}} \color[HTML]{000000} 0.043 & {\cellcolor[HTML]{180C0A}} \color{white} -1.132 & {\cellcolor[HTML]{2D0E00}} \color{white} -0.137 & {\cellcolor[HTML]{170D0B}} \color{white} 1.057 \\
bm25+reranker+weighted [tf-idk] & {\cellcolor[HTML]{4997C9}} \color[HTML]{F1F1F1} 0.176 & {\cellcolor[HTML]{4695C8}} \color[HTML]{F1F1F1} 0.123 & {\cellcolor[HTML]{180C0A}} \color{white} -1.482 & {\cellcolor[HTML]{843924}} \color{white} -0.494 & {\cellcolor[HTML]{3B1100}} \color{white} -0.048 \\
B3-SBERT-V1-TestAbs [sambs] & {\cellcolor[HTML]{CFE1F2}} \color[HTML]{000000} 0.060 & {\cellcolor[HTML]{CEE0F2}} \color[HTML]{000000} 0.042 & {\cellcolor[HTML]{180C0A}} \color{white} -1.517 & {\cellcolor[HTML]{340F00}} \color{white} -0.170 & {\cellcolor[HTML]{170D0B}} \color{white} 1.051 \\
bm25+reranker [tf-idk] & {\cellcolor[HTML]{4191C6}} \color[HTML]{F1F1F1} 0.183 & {\cellcolor[HTML]{4191C6}} \color[HTML]{F1F1F1} 0.126 & {\cellcolor[HTML]{180C0A}} \color{white} -1.663 & {\cellcolor[HTML]{944834}} \color{white} -0.553 & {\cellcolor[HTML]{3F1201}} \color{white} -0.165 \\
BM25\_k1\_0p95\_b\_0p75\_stop\_stem\_fullText [academy-retrievals] & {\cellcolor[HTML]{08316D}} \color[HTML]{F1F1F1} 0.290 & {\cellcolor[HTML]{083573}} \color[HTML]{F1F1F1} 0.197 & {\cellcolor[HTML]{180C0A}} \color{white} -1.817 & {\cellcolor[HTML]{FBA9A8}} \color{black} -0.920 & {\cellcolor[HTML]{FFADAD}} \color{black} -3.696 \\
BM25\_k1\_0p2\_b\_0p8\_stop\_stem\_fullText [academy-retrievals] & {\cellcolor[HTML]{08306B}} \color[HTML]{F1F1F1} 0.291 & {\cellcolor[HTML]{083674}} \color[HTML]{F1F1F1} 0.196 & {\cellcolor[HTML]{180C0A}} \color{white} -1.862 & {\cellcolor[HTML]{FFADAD}} \color{black} -0.935 & {\cellcolor[HTML]{F9A7A6}} \color{black} -3.569 \\
BM25\_k1\_1p0\_b\_0p7\_stop\_stem\_fullText [academy-retrievals] & {\cellcolor[HTML]{08306B}} \color[HTML]{F1F1F1} 0.292 & {\cellcolor[HTML]{083573}} \color[HTML]{F1F1F1} 0.197 & {\cellcolor[HTML]{180C0A}} \color{white} -1.865 & {\cellcolor[HTML]{FFADAD}} \color{black} -0.939 & {\cellcolor[HTML]{FDABAB}} \color{black} -3.628 \\
B4-SBERT-V2-MPNet-TestFull [sambs] & {\cellcolor[HTML]{CADDF0}} \color[HTML]{000000} 0.068 & {\cellcolor[HTML]{CADEF0}} \color[HTML]{000000} 0.046 & {\cellcolor[HTML]{180C0A}} \color{white} -1.867 & {\cellcolor[HTML]{3C1101}} \color{white} -0.218 & {\cellcolor[HTML]{190C09}} \color{white} 0.978 \\
B3-SBERT-V2-MPNet-TestAbs [sambs] & {\cellcolor[HTML]{CADEF0}} \color[HTML]{000000} 0.067 & {\cellcolor[HTML]{CBDEF1}} \color[HTML]{000000} 0.045 & {\cellcolor[HTML]{180C0A}} \color{white} -1.892 & {\cellcolor[HTML]{3C1101}} \color{white} -0.215 & {\cellcolor[HTML]{190C09}} \color{white} 0.983 \\
BM25\_ColBERTv2\_ReRanker\_v1 [academy-retrievals] & {\cellcolor[HTML]{4896C8}} \color[HTML]{F1F1F1} 0.177 & {\cellcolor[HTML]{4D99CA}} \color[HTML]{F1F1F1} 0.119 & {\cellcolor[HTML]{180C0A}} \color{white} -1.916 & {\cellcolor[HTML]{9A4D3B}} \color{white} -0.575 & {\cellcolor[HTML]{3B1100}} \color{white} -0.031 \\
qrel-boost-core \cite{OWS} & {\cellcolor[HTML]{1966AD}} \color[HTML]{F1F1F1} 0.231 & {\cellcolor[HTML]{2777B8}} \color[HTML]{F1F1F1} 0.146 & {\cellcolor[HTML]{180C0A}} \color{white} -2.789 & {\cellcolor[HTML]{F2A19F}} \color{black} -0.888 & {\cellcolor[HTML]{5D1E09}} \color{white} -0.800 \\
query-intent-fusion \cite{OWS} & {\cellcolor[HTML]{1966AD}} \color[HTML]{F1F1F1} 0.231 & {\cellcolor[HTML]{2777B8}} \color[HTML]{F1F1F1} 0.146 & {\cellcolor[HTML]{180C0A}} \color{white} -2.789 & {\cellcolor[HTML]{F2A19F}} \color{black} -0.888 & {\cellcolor[HTML]{5D1E09}} \color{white} -0.800 \\
\bottomrule
\end{tabular}
}
\end{table}

The ER describes the extent to which the system effect is recovered. If the same system effect is achieved in the evolved setting, the ER is 1, which describes a perfectly persistent system. The approaches \texttt{ci}, \texttt{dsgt-bm25-submission}, and \texttt{ps} show the highest recovery of the system effect over the short term. For the long-term change setting, the scores diverge strongly. \texttt{seupd2425-rand-DL-run}, \texttt{seupd2425-rand-elision+synonyms0.5}, and \texttt{bm25-baseline+traditional} have an ER closest to 1.

The results for the Sci task are presented in Table~\ref{tab:sci-results-change}. The systems that ranked too few documents and the BM25 reference approach were excluded from this evaluation.
The change measured by RI and DRI does not necessarily agree with each other. The approaches with the lowest change according the RI are \texttt{fusion-with-core}, \texttt{monot5-in-core}, and \texttt{ows-bm25}. The DRI states \texttt{monot5-in-core}, \texttt{monot5-in-core}, \texttt{ows-bm25}, and \texttt{ows-cluster-boosting}. Besides the \texttt{fusion-with-core}, all approaches improve in effectiveness. This approach and additionally \texttt{monot5-in-core} do not show an improved system effect while the difference for \texttt{monot5-in-core} is only marginal.

Regarding the ER, the systems \texttt{B4-SBERT-V2-MPNet-TestFull}, \texttt{B3-SBERT-{\allowbreak}V2-{\allowbreak}MPNet-TestAbs}, and \texttt{BM25\_ColBERTv2\_ReRanker\_v1} have an ER closest to 1, indicating the highest persistence. The measure seems not to agree with the change-based measures described before. 

\begin{table}[htb!]
\caption{Short-term change results for the WebRetrieval task in comparison to the snapshot 2023-03. The results are sorted by RI for the 2023-05 snapshot. Darker colors indicate less change or higher effectiveness.}
\label{tab:web-results-change-short}
\resizebox{\textwidth}{!}{%

\begin{tabular}{lc@{\hskip 5pt}c@{\hskip 5pt}c@{\hskip 5pt}c@{\hskip 5pt}c@{\hskip 5pt}c}
\toprule
Approach & ARP & MARP & RI & DRI & ER \\
\midrule
run2 \cite{RACOON} & {\cellcolor[HTML]{2070B4}} \color[HTML]{F1F1F1} 0.385 & {\cellcolor[HTML]{115CA5}} \color[HTML]{F1F1F1} 0.436 & {\cellcolor[HTML]{9EB0FF}} \color{black} 0.210 & {\cellcolor[HTML]{9EB0FF}} \color{black} 0.400 & {\cellcolor[HTML]{863B26}} \color{white} 0.448 \\
run3 \cite{RACOON} & {\cellcolor[HTML]{1C6BB0}} \color[HTML]{F1F1F1} 0.394 & {\cellcolor[HTML]{105BA4}} \color[HTML]{F1F1F1} 0.438 & {\cellcolor[HTML]{76ABEC}} \color{black} 0.182 & {\cellcolor[HTML]{7BACEE}} \color{black} 0.351 & {\cellcolor[HTML]{752D17}} \color{white} 0.507 \\
run1 \cite{RACOON} & {\cellcolor[HTML]{084A91}} \color[HTML]{F1F1F1} 0.455 & {\cellcolor[HTML]{083877}} \color[HTML]{F1F1F1} 0.501 & {\cellcolor[HTML]{62A6E0}} \color{black} 0.168 & {\cellcolor[HTML]{8CAEF6}} \color{black} 0.372 & {\cellcolor[HTML]{591C07}} \color{white} 0.613 \\
JMFT\_Bert \cite{CIR} & {\cellcolor[HTML]{2777B8}} \color[HTML]{F1F1F1} 0.372 & {\cellcolor[HTML]{2474B7}} \color[HTML]{F1F1F1} 0.390 & {\cellcolor[HTML]{20546C}} \color{white} 0.086 & {\cellcolor[HTML]{20546C}} \color{white} 0.164 & {\cellcolor[HTML]{541905}} \color{white} 0.637 \\
query-variants-qrel-boost-kmeans \cite{CIRcluster} & {\cellcolor[HTML]{084E98}} \color[HTML]{F1F1F1} 0.447 & {\cellcolor[HTML]{084B93}} \color[HTML]{F1F1F1} 0.467 & {\cellcolor[HTML]{1F5068}} \color{white} 0.083 & {\cellcolor[HTML]{276481}} \color{white} 0.191 & {\cellcolor[HTML]{3E1201}} \color{white} 0.755 \\
JMFT\_Jaccard \cite{CIR} & {\cellcolor[HTML]{2676B8}} \color[HTML]{F1F1F1} 0.375 & {\cellcolor[HTML]{2474B7}} \color[HTML]{F1F1F1} 0.391 & {\cellcolor[HTML]{1E4D63}} \color{white} 0.079 & {\cellcolor[HTML]{1E4E65}} \color{white} 0.154 & {\cellcolor[HTML]{4F1703}} \color{white} 0.661 \\
JMFT\_lenght \cite{CIR} & {\cellcolor[HTML]{2676B8}} \color[HTML]{F1F1F1} 0.375 & {\cellcolor[HTML]{2474B7}} \color[HTML]{F1F1F1} 0.391 & {\cellcolor[HTML]{1E4D63}} \color{white} 0.079 & {\cellcolor[HTML]{1E4E65}} \color{white} 0.154 & {\cellcolor[HTML]{4F1703}} \color{white} 0.661 \\
query-variants-relevance-feedback-kmeans \cite{CIRcluster} & {\cellcolor[HTML]{6DAFD7}} \color[HTML]{F1F1F1} 0.265 & {\cellcolor[HTML]{6CAED6}} \color[HTML]{F1F1F1} 0.275 & {\cellcolor[HTML]{183D4F}} \color{white} 0.064 & {\cellcolor[HTML]{132D3A}} \color{white} 0.093 & {\cellcolor[HTML]{9EB0FF}} \color{black} 18.201 \\
baseline-qrel-boost \cite{CIRcluster} & {\cellcolor[HTML]{08306B}} \color[HTML]{F1F1F1} 0.501 & {\cellcolor[HTML]{08306B}} \color[HTML]{F1F1F1} 0.518 & {\cellcolor[HTML]{173C4D}} \color{white} 0.063 & {\cellcolor[HTML]{225973}} \color{white} 0.173 & {\cellcolor[HTML]{310F00}} \color{white} 0.828 \\
query-variants-qrel-boost-dbscan \cite{CIRcluster} & {\cellcolor[HTML]{1967AD}} \color[HTML]{F1F1F1} 0.402 & {\cellcolor[HTML]{1A68AE}} \color[HTML]{F1F1F1} 0.414 & {\cellcolor[HTML]{153544}} \color{white} 0.057 & {\cellcolor[HTML]{194153}} \color{white} 0.130 & {\cellcolor[HTML]{3C1101}} \color{white} 0.762 \\
query-variants-qrel-boost \cite{CIRcluster} & {\cellcolor[HTML]{1967AD}} \color[HTML]{F1F1F1} 0.402 & {\cellcolor[HTML]{1A68AE}} \color[HTML]{F1F1F1} 0.414 & {\cellcolor[HTML]{153544}} \color{white} 0.057 & {\cellcolor[HTML]{194153}} \color{white} 0.130 & {\cellcolor[HTML]{3C1101}} \color{white} 0.762 \\
baseline-relevance-feedback \cite{CIRcluster} & {\cellcolor[HTML]{0F5AA3}} \color[HTML]{F1F1F1} 0.426 & {\cellcolor[HTML]{135FA7}} \color[HTML]{F1F1F1} 0.430 & {\cellcolor[HTML]{11161B}} \color{white} 0.021 & {\cellcolor[HTML]{112630}} \color{white} 0.078 & {\cellcolor[HTML]{2A0E01}} \color{white} 0.878 \\
dense-baseline+multilingual [air5] & {\cellcolor[HTML]{F7FBFF}} \color[HTML]{000000} 0.034 & {\cellcolor[HTML]{F7FBFF}} \color[HTML]{000000} 0.034 & {\cellcolor[HTML]{111519}} \color{white} 0.019 & {\cellcolor[HTML]{170D0B}} \color{white} 0.006 & {\cellcolor[HTML]{180C0A}} \color{white} 1.039 \\
seupd-2425-basette \cite{BASETTE} & {\cellcolor[HTML]{63A8D3}} \color[HTML]{F1F1F1} 0.279 & {\cellcolor[HTML]{68ACD5}} \color[HTML]{F1F1F1} 0.281 & {\cellcolor[HTML]{121214}} \color{white} 0.013 & {\cellcolor[HTML]{11181C}} \color{white} 0.043 & {\cellcolor[HTML]{153342}} \color{white} 5.459 \\
query-variants-relevance-feedback-dbscan \cite{CIRcluster} & {\cellcolor[HTML]{519CCC}} \color[HTML]{F1F1F1} 0.305 & {\cellcolor[HTML]{57A0CE}} \color[HTML]{F1F1F1} 0.306 & {\cellcolor[HTML]{140F10}} \color{white} 0.008 & {\cellcolor[HTML]{11181C}} \color{white} 0.042 & {\cellcolor[HTML]{803620}} \color{white} 0.470 \\
bm25-baseline+traditional [air5] & {\cellcolor[HTML]{B9D6EA}} \color[HTML]{000000} 0.170 & {\cellcolor[HTML]{BDD7EC}} \color[HTML]{000000} 0.170 & {\cellcolor[HTML]{140F10}} \color{white} 0.007 & {\cellcolor[HTML]{121214}} \color{white} 0.023 & {\cellcolor[HTML]{180C0A}} \color{white} 1.105 \\
clef25-seupd2425-rise \cite{RISE} & {\cellcolor[HTML]{083877}} \color[HTML]{F1F1F1} 0.485 & {\cellcolor[HTML]{084184}} \color[HTML]{F1F1F1} 0.487 & {\cellcolor[HTML]{140F10}} \color{white} 0.007 & {\cellcolor[HTML]{112028}} \color{white} 0.064 & {\cellcolor[HTML]{210B03}} \color{white} 0.939 \\
datahunter-web-20250430-run1 \cite{DataHunter} & {\cellcolor[HTML]{4191C6}} \color[HTML]{F1F1F1} 0.326 & {\cellcolor[HTML]{4997C9}} \color[HTML]{F1F1F1} 0.327 & {\cellcolor[HTML]{150E0E}} \color{white} 0.006 & {\cellcolor[HTML]{11181C}} \color{white} 0.043 & {\cellcolor[HTML]{451401}} \color{white} 0.709 \\
fair\_schaer \cite{CIR} & {\cellcolor[HTML]{65AAD4}} \color[HTML]{F1F1F1} 0.276 & {\cellcolor[HTML]{6AAED6}} \color[HTML]{F1F1F1} 0.277 & {\cellcolor[HTML]{150E0E}} \color{white} 0.006 & {\cellcolor[HTML]{111519}} \color{white} 0.036 & {\cellcolor[HTML]{111317}} \color{white} 2.326 \\
molten-marsanne \cite{CIR} & {\cellcolor[HTML]{4191C6}} \color[HTML]{F1F1F1} 0.327 & {\cellcolor[HTML]{4896C8}} \color[HTML]{F1F1F1} 0.328 & {\cellcolor[HTML]{160E0D}} \color{white} 0.004 & {\cellcolor[HTML]{11161B}} \color{white} 0.040 & {\cellcolor[HTML]{3E1201}} \color{white} 0.750 \\
seupd2425-rand-ICU-run \cite{RAND} & {\cellcolor[HTML]{95C5DF}} \color[HTML]{000000} 0.220 & {\cellcolor[HTML]{9AC8E0}} \color[HTML]{000000} 0.220 & {\cellcolor[HTML]{1A0C08}} \color{white} -0.001 & {\cellcolor[HTML]{121214}} \color{white} 0.023 & {\cellcolor[HTML]{170D0B}} \color{white} 1.145 \\
seupd2425-rand-elision+synonyms0.5 \cite{RAND} & {\cellcolor[HTML]{A6CEE4}} \color[HTML]{000000} 0.198 & {\cellcolor[HTML]{AACFE5}} \color[HTML]{000000} 0.198 & {\cellcolor[HTML]{1B0B07}} \color{white} -0.002 & {\cellcolor[HTML]{131112}} \color{white} 0.020 & {\cellcolor[HTML]{180C0A}} \color{white} 1.106 \\
dsgt-bm25-rerank-submission \cite{DS@GT} & {\cellcolor[HTML]{135FA7}} \color[HTML]{F1F1F1} 0.416 & {\cellcolor[HTML]{1967AD}} \color[HTML]{F1F1F1} 0.415 & {\cellcolor[HTML]{1C0B06}} \color{white} -0.003 & {\cellcolor[HTML]{11181C}} \color{white} 0.042 & {\cellcolor[HTML]{210B03}} \color{white} 0.936 \\
seupd2425-sard-minAnaBool1000-GPT \cite{SARD} & {\cellcolor[HTML]{4D99CA}} \color[HTML]{F1F1F1} 0.311 & {\cellcolor[HTML]{549FCD}} \color[HTML]{F1F1F1} 0.310 & {\cellcolor[HTML]{1C0B06}} \color{white} -0.003 & {\cellcolor[HTML]{111317}} \color{white} 0.031 & {\cellcolor[HTML]{541905}} \color{white} 0.643 \\
seupd2425-sard-minAnaPhr300 \cite{SARD} & {\cellcolor[HTML]{3080BD}} \color[HTML]{F1F1F1} 0.358 & {\cellcolor[HTML]{3787C0}} \color[HTML]{F1F1F1} 0.357 & {\cellcolor[HTML]{1E0B04}} \color{white} -0.005 & {\cellcolor[HTML]{111418}} \color{white} 0.033 & {\cellcolor[HTML]{270D01}} \color{white} 0.891 \\
seupd2425-sard-minAnaMixed3-GPT \cite{SARD} & {\cellcolor[HTML]{206FB4}} \color[HTML]{F1F1F1} 0.387 & {\cellcolor[HTML]{2777B8}} \color[HTML]{F1F1F1} 0.385 & {\cellcolor[HTML]{250C01}} \color{white} -0.010 & {\cellcolor[HTML]{111317}} \color{white} 0.029 & {\cellcolor[HTML]{210B03}} \color{white} 0.943 \\
seupd2425-rand-frenchfilter-run \cite{RAND} & {\cellcolor[HTML]{9DCAE1}} \color[HTML]{000000} 0.210 & {\cellcolor[HTML]{A3CCE3}} \color[HTML]{000000} 0.209 & {\cellcolor[HTML]{250C01}} \color{white} -0.010 & {\cellcolor[HTML]{140F10}} \color{white} 0.015 & {\cellcolor[HTML]{180C0A}} \color{white} 1.097 \\
seupd2425-sard-minAnaMixed \cite{SARD} & {\cellcolor[HTML]{2474B7}} \color[HTML]{F1F1F1} 0.378 & {\cellcolor[HTML]{2D7DBB}} \color[HTML]{F1F1F1} 0.375 & {\cellcolor[HTML]{270D01}} \color{white} -0.011 & {\cellcolor[HTML]{121215}} \color{white} 0.026 & {\cellcolor[HTML]{210B03}} \color{white} 0.939 \\
rr-ps \cite{3DS2A} & {\cellcolor[HTML]{1460A8}} \color[HTML]{F1F1F1} 0.415 & {\cellcolor[HTML]{1B69AF}} \color[HTML]{F1F1F1} 0.412 & {\cellcolor[HTML]{2B0E01}} \color{white} -0.014 & {\cellcolor[HTML]{121215}} \color{white} 0.025 & {\cellcolor[HTML]{1C0B06}} \color{white} 0.970 \\
run4 \cite{RACOON} & {\cellcolor[HTML]{2070B4}} \color[HTML]{F1F1F1} 0.385 & {\cellcolor[HTML]{2A7AB9}} \color[HTML]{F1F1F1} 0.381 & {\cellcolor[HTML]{310F00}} \color{white} -0.018 & {\cellcolor[HTML]{141011}} \color{white} 0.018 & {\cellcolor[HTML]{1C0B06}} \color{white} 0.975 \\
ps \cite{3DS2A} & {\cellcolor[HTML]{1865AC}} \color[HTML]{F1F1F1} 0.406 & {\cellcolor[HTML]{1F6EB3}} \color[HTML]{F1F1F1} 0.402 & {\cellcolor[HTML]{330F00}} \color{white} -0.019 & {\cellcolor[HTML]{141011}} \color{white} 0.018 & {\cellcolor[HTML]{1A0C08}} \color{white} 0.985 \\
seupd2425-rand-DL-run \cite{RAND} & {\cellcolor[HTML]{A1CBE2}} \color[HTML]{000000} 0.205 & {\cellcolor[HTML]{A6CEE4}} \color[HTML]{000000} 0.203 & {\cellcolor[HTML]{340F00}} \color{white} -0.020 & {\cellcolor[HTML]{160E0D}} \color{white} 0.008 & {\cellcolor[HTML]{180C0A}} \color{white} 1.068 \\
query\_expansion\_time\_dependence \cite{CIR} & {\cellcolor[HTML]{58A1CF}} \color[HTML]{F1F1F1} 0.294 & {\cellcolor[HTML]{61A7D2}} \color[HTML]{F1F1F1} 0.291 & {\cellcolor[HTML]{391100}} \color{white} -0.023 & {\cellcolor[HTML]{160E0D}} \color{white} 0.009 & {\cellcolor[HTML]{FFADAD}} \color{black} -0.035 \\
ci \cite{3DS2A} & {\cellcolor[HTML]{2C7CBA}} \color[HTML]{F1F1F1} 0.366 & {\cellcolor[HTML]{3484BF}} \color[HTML]{F1F1F1} 0.362 & {\cellcolor[HTML]{3B1100}} \color{white} -0.024 & {\cellcolor[HTML]{150E0E}} \color{white} 0.010 & {\cellcolor[HTML]{1A0C08}} \color{white} 0.989 \\
mein-cleveres-system \cite{CIR} & {\cellcolor[HTML]{58A1CF}} \color[HTML]{F1F1F1} 0.294 & {\cellcolor[HTML]{61A7D2}} \color[HTML]{F1F1F1} 0.291 & {\cellcolor[HTML]{3B1100}} \color{white} -0.024 & {\cellcolor[HTML]{160E0D}} \color{white} 0.008 & {\cellcolor[HTML]{F6A5A3}} \color{black} 0.000 \\
dsgt-bm25-submission \cite{DS@GT} & {\cellcolor[HTML]{3787C0}} \color[HTML]{F1F1F1} 0.345 & {\cellcolor[HTML]{4090C5}} \color[HTML]{F1F1F1} 0.340 & {\cellcolor[HTML]{3C1101}} \color{white} -0.025 & {\cellcolor[HTML]{160E0D}} \color{white} 0.007 & {\cellcolor[HTML]{1A0C08}} \color{white} 0.986 \\
clef-webbusters-traditional [web-busters] & {\cellcolor[HTML]{5AA2CF}} \color[HTML]{F1F1F1} 0.292 & {\cellcolor[HTML]{63A8D3}} \color[HTML]{F1F1F1} 0.288 & {\cellcolor[HTML]{3E1201}} \color{white} -0.026 & {\cellcolor[HTML]{170D0B}} \color{white} 0.006 & {\cellcolor[HTML]{112630}} \color{white} 4.277 \\
searchill\_YYYY-MM [searchill] & {\cellcolor[HTML]{4F9BCB}} \color[HTML]{F1F1F1} 0.306 & {\cellcolor[HTML]{5AA2CF}} \color[HTML]{F1F1F1} 0.302 & {\cellcolor[HTML]{451401}} \color{white} -0.030 & {\cellcolor[HTML]{180C0A}} \color{white} 0.002 & {\cellcolor[HTML]{1C0B06}} \color{white} 0.973 \\
ps-prf \cite{3DS2A} & {\cellcolor[HTML]{3282BE}} \color[HTML]{F1F1F1} 0.354 & {\cellcolor[HTML]{3B8BC2}} \color[HTML]{F1F1F1} 0.349 & {\cellcolor[HTML]{471401}} \color{white} -0.031 & {\cellcolor[HTML]{180C0A}} \color{white} 0.001 & {\cellcolor[HTML]{180C0A}} \color{white} 1.025 \\
ci-prf-ps \cite{3DS2A} & {\cellcolor[HTML]{2D7DBB}} \color[HTML]{F1F1F1} 0.364 & {\cellcolor[HTML]{3686C0}} \color[HTML]{F1F1F1} 0.358 & {\cellcolor[HTML]{4B1602}} \color{white} -0.033 & {\cellcolor[HTML]{1C0B06}} \color{white} -0.001 & {\cellcolor[HTML]{180C0A}} \color{white} 1.034 \\
bm25\_boosted\_temporal\_categories \cite{CIR} & {\cellcolor[HTML]{8ABFDD}} \color[HTML]{000000} 0.232 & {\cellcolor[HTML]{95C5DF}} \color[HTML]{000000} 0.226 & {\cellcolor[HTML]{7B321C}} \color{white} -0.054 & {\cellcolor[HTML]{65230E}} \color{white} -0.016 & {\cellcolor[HTML]{1E0B04}} \color{white} 0.955 \\
luyuHF [air5] & {\cellcolor[HTML]{BDD7EC}} \color[HTML]{000000} 0.165 & {\cellcolor[HTML]{C4DAEE}} \color[HTML]{000000} 0.157 & {\cellcolor[HTML]{FFADAD}} \color{black} -0.108 & {\cellcolor[HTML]{FFADAD}} \color{black} -0.038 & {\cellcolor[HTML]{180C0A}} \color{white} 1.049 \\
\bottomrule
\end{tabular}
}
\end{table}

\begin{table}[htb!]
\caption{Long-term change results for the WebRetrieval task in comparison to the snapshot 2023-03. The results are sorted by RI for the 2023-08 snapshot. Darker colors indicate less change or higher effectiveness.}
\label{tab:web-results-change-long}
\resizebox{\textwidth}{!}{%

\begin{tabular}{lc@{\hskip 5pt}c@{\hskip 5pt}c@{\hskip 5pt}c@{\hskip 5pt}c@{\hskip 5pt}c}
\toprule
Approach & ARP & MARP & RI & DRI & ER \\
\midrule
run2 \cite{RACOON} & {\cellcolor[HTML]{1865AC}} \color[HTML]{F1F1F1} 0.320 & {\cellcolor[HTML]{0A549E}} \color[HTML]{F1F1F1} 0.403 & {\cellcolor[HTML]{9EB0FF}} \color{black} 0.344 & {\cellcolor[HTML]{9AB0FD}} \color{black} 0.456 & {\cellcolor[HTML]{6E2813}} \color{white} 0.318 \\
run1 \cite{RACOON} & {\cellcolor[HTML]{083979}} \color[HTML]{F1F1F1} 0.371 & {\cellcolor[HTML]{08306B}} \color[HTML]{F1F1F1} 0.459 & {\cellcolor[HTML]{8CAEF6}} \color{black} 0.322 & {\cellcolor[HTML]{9EB0FF}} \color{black} 0.466 & {\cellcolor[HTML]{591C07}} \color{white} 0.440 \\
run3 \cite{RACOON} & {\cellcolor[HTML]{125EA6}} \color[HTML]{F1F1F1} 0.328 & {\cellcolor[HTML]{0A539E}} \color[HTML]{F1F1F1} 0.405 & {\cellcolor[HTML]{8CAEF6}} \color{black} 0.320 & {\cellcolor[HTML]{79ABED}} \color{black} 0.407 & {\cellcolor[HTML]{65230E}} \color{white} 0.369 \\
query-variants-qrel-boost-kmeans \cite{CIRcluster} & {\cellcolor[HTML]{09529D}} \color[HTML]{F1F1F1} 0.342 & {\cellcolor[HTML]{084C95}} \color[HTML]{F1F1F1} 0.415 & {\cellcolor[HTML]{76ABEC}} \color{black} 0.298 & {\cellcolor[HTML]{60A5DF}} \color{black} 0.370 & {\cellcolor[HTML]{5B1D08}} \color{white} 0.429 \\
baseline-qrel-boost \cite{CIRcluster} & {\cellcolor[HTML]{083573}} \color[HTML]{F1F1F1} 0.376 & {\cellcolor[HTML]{08316D}} \color[HTML]{F1F1F1} 0.456 & {\cellcolor[HTML]{76ABEC}} \color{black} 0.297 & {\cellcolor[HTML]{76ABEC}} \color{black} 0.402 & {\cellcolor[HTML]{521804}} \color{white} 0.484 \\
query-variants-qrel-boost \cite{CIRcluster} & {\cellcolor[HTML]{1F6EB3}} \color[HTML]{F1F1F1} 0.309 & {\cellcolor[HTML]{1C6AB0}} \color[HTML]{F1F1F1} 0.368 & {\cellcolor[HTML]{62A6E0}} \color{black} 0.276 & {\cellcolor[HTML]{3584AA}} \color{white} 0.286 & {\cellcolor[HTML]{63210C}} \color{white} 0.379 \\
query-variants-qrel-boost-dbscan \cite{CIRcluster} & {\cellcolor[HTML]{1F6EB3}} \color[HTML]{F1F1F1} 0.309 & {\cellcolor[HTML]{1C6AB0}} \color[HTML]{F1F1F1} 0.368 & {\cellcolor[HTML]{62A6E0}} \color{black} 0.276 & {\cellcolor[HTML]{3584AA}} \color{white} 0.286 & {\cellcolor[HTML]{63210C}} \color{white} 0.379 \\
JMFT\_Jaccard \cite{CIR} & {\cellcolor[HTML]{2676B8}} \color[HTML]{F1F1F1} 0.299 & {\cellcolor[HTML]{2474B7}} \color[HTML]{F1F1F1} 0.353 & {\cellcolor[HTML]{5BA4DB}} \color{white} 0.267 & {\cellcolor[HTML]{2E7699}} \color{white} 0.259 & {\cellcolor[HTML]{68240F}} \color{white} 0.353 \\
baseline-relevance-feedback \cite{CIRcluster} & {\cellcolor[HTML]{1865AC}} \color[HTML]{F1F1F1} 0.319 & {\cellcolor[HTML]{1764AB}} \color[HTML]{F1F1F1} 0.377 & {\cellcolor[HTML]{58A3D9}} \color{white} 0.266 & {\cellcolor[HTML]{327EA3}} \color{white} 0.275 & {\cellcolor[HTML]{5B1D08}} \color{white} 0.425 \\
JMFT\_Bert \cite{CIR} & {\cellcolor[HTML]{2474B7}} \color[HTML]{F1F1F1} 0.301 & {\cellcolor[HTML]{2373B6}} \color[HTML]{F1F1F1} 0.354 & {\cellcolor[HTML]{54A0D5}} \color{white} 0.260 & {\cellcolor[HTML]{2C7192}} \color{white} 0.249 & {\cellcolor[HTML]{63210C}} \color{white} 0.374 \\
JMFT\_lenght \cite{CIR} & {\cellcolor[HTML]{2474B7}} \color[HTML]{F1F1F1} 0.301 & {\cellcolor[HTML]{2373B6}} \color[HTML]{F1F1F1} 0.354 & {\cellcolor[HTML]{54A0D5}} \color{white} 0.260 & {\cellcolor[HTML]{2C7192}} \color{white} 0.249 & {\cellcolor[HTML]{63210C}} \color{white} 0.374 \\
query-variants-relevance-feedback-kmeans \cite{CIRcluster} & {\cellcolor[HTML]{7AB6D9}} \color[HTML]{000000} 0.217 & {\cellcolor[HTML]{72B2D8}} \color[HTML]{F1F1F1} 0.250 & {\cellcolor[HTML]{4194C1}} \color{white} 0.235 & {\cellcolor[HTML]{183D4F}} \color{white} 0.145 & {\cellcolor[HTML]{9EB0FF}} \color{black} 24.220 \\
clef25-seupd2425-rise \cite{RISE} & {\cellcolor[HTML]{08306B}} \color[HTML]{F1F1F1} 0.383 & {\cellcolor[HTML]{083E81}} \color[HTML]{F1F1F1} 0.436 & {\cellcolor[HTML]{3787AF}} \color{white} 0.217 & {\cellcolor[HTML]{25607C}} \color{white} 0.215 & {\cellcolor[HTML]{3F1201}} \color{white} 0.626 \\
query-variants-relevance-feedback-dbscan \cite{CIRcluster} & {\cellcolor[HTML]{5AA2CF}} \color[HTML]{F1F1F1} 0.244 & {\cellcolor[HTML]{5BA3D0}} \color[HTML]{F1F1F1} 0.276 & {\cellcolor[HTML]{3280A6}} \color{white} 0.205 & {\cellcolor[HTML]{153342}} \color{white} 0.121 & {\cellcolor[HTML]{FFADAD}} \color{black} -0.522 \\
seupd-2425-basette \cite{BASETTE} & {\cellcolor[HTML]{6AAED6}} \color[HTML]{F1F1F1} 0.228 & {\cellcolor[HTML]{6CAED6}} \color[HTML]{F1F1F1} 0.255 & {\cellcolor[HTML]{2F789C}} \color{white} 0.196 & {\cellcolor[HTML]{122A36}} \color{white} 0.101 & {\cellcolor[HTML]{1E4D63}} \color{white} 9.776 \\
dsgt-bm25-rerank-submission \cite{DS@GT} & {\cellcolor[HTML]{0C56A0}} \color[HTML]{F1F1F1} 0.337 & {\cellcolor[HTML]{1865AC}} \color[HTML]{F1F1F1} 0.376 & {\cellcolor[HTML]{2C7394}} \color{white} 0.188 & {\cellcolor[HTML]{173A4B}} \color{white} 0.137 & {\cellcolor[HTML]{3F1201}} \color{white} 0.628 \\
datahunter-web-20250430-run1 \cite{DataHunter} & {\cellcolor[HTML]{4191C6}} \color[HTML]{F1F1F1} 0.267 & {\cellcolor[HTML]{4997C9}} \color[HTML]{F1F1F1} 0.298 & {\cellcolor[HTML]{2C7394}} \color{white} 0.188 & {\cellcolor[HTML]{132D3A}} \color{white} 0.107 & {\cellcolor[HTML]{803620}} \color{white} 0.217 \\
seupd2425-sard-minAnaBool1000-GPT \cite{SARD} & {\cellcolor[HTML]{529DCC}} \color[HTML]{F1F1F1} 0.252 & {\cellcolor[HTML]{57A0CE}} \color[HTML]{F1F1F1} 0.281 & {\cellcolor[HTML]{2C7394}} \color{white} 0.187 & {\cellcolor[HTML]{122A36}} \color{white} 0.100 & {\cellcolor[HTML]{BC6D61}} \color{white} -0.130 \\
seupd2425-rand-ICU-run \cite{RAND} & {\cellcolor[HTML]{A8CEE4}} \color[HTML]{000000} 0.180 & {\cellcolor[HTML]{A3CCE3}} \color[HTML]{000000} 0.200 & {\cellcolor[HTML]{2B6F8F}} \color{white} 0.182 & {\cellcolor[HTML]{101D25}} \color{white} 0.067 & {\cellcolor[HTML]{180C0A}} \color{white} 1.177 \\
molten-marsanne \cite{CIR} & {\cellcolor[HTML]{3F8FC5}} \color[HTML]{F1F1F1} 0.270 & {\cellcolor[HTML]{4997C9}} \color[HTML]{F1F1F1} 0.299 & {\cellcolor[HTML]{2A6D8D}} \color{white} 0.179 & {\cellcolor[HTML]{122934}} \color{white} 0.096 & {\cellcolor[HTML]{6C2711}} \color{white} 0.325 \\
seupd2425-sard-minAnaMixed3-GPT \cite{SARD} & {\cellcolor[HTML]{1B69AF}} \color[HTML]{F1F1F1} 0.315 & {\cellcolor[HTML]{2676B8}} \color[HTML]{F1F1F1} 0.349 & {\cellcolor[HTML]{296B8B}} \color{white} 0.177 & {\cellcolor[HTML]{132D3A}} \color{white} 0.109 & {\cellcolor[HTML]{411201}} \color{white} 0.611 \\
seupd2425-sard-minAnaMixed \cite{SARD} & {\cellcolor[HTML]{1E6DB2}} \color[HTML]{F1F1F1} 0.310 & {\cellcolor[HTML]{2C7CBA}} \color[HTML]{F1F1F1} 0.341 & {\cellcolor[HTML]{286886}} \color{white} 0.171 & {\cellcolor[HTML]{122934}} \color{white} 0.097 & {\cellcolor[HTML]{411201}} \color{white} 0.614 \\
rr-ps \cite{3DS2A} & {\cellcolor[HTML]{0A549E}} \color[HTML]{F1F1F1} 0.340 & {\cellcolor[HTML]{1966AD}} \color[HTML]{F1F1F1} 0.375 & {\cellcolor[HTML]{276683}} \color{white} 0.169 & {\cellcolor[HTML]{122C38}} \color{white} 0.103 & {\cellcolor[HTML]{381000}} \color{white} 0.683 \\
fair\_schaer \cite{CIR} & {\cellcolor[HTML]{68ACD5}} \color[HTML]{F1F1F1} 0.231 & {\cellcolor[HTML]{6CAED6}} \color[HTML]{F1F1F1} 0.255 & {\cellcolor[HTML]{276683}} \color{white} 0.169 & {\cellcolor[HTML]{101F26}} \color{white} 0.070 & {\cellcolor[HTML]{111519}} \color{white} 3.130 \\
seupd2425-rand-elision+synonyms0.5 \cite{RAND} & {\cellcolor[HTML]{B8D5EA}} \color[HTML]{000000} 0.165 & {\cellcolor[HTML]{B2D2E8}} \color[HTML]{000000} 0.181 & {\cellcolor[HTML]{276481}} \color{white} 0.166 & {\cellcolor[HTML]{11181C}} \color{white} 0.048 & {\cellcolor[HTML]{180C0A}} \color{white} 1.055 \\
seupd2425-rand-frenchfilter-run \cite{RAND} & {\cellcolor[HTML]{AED1E7}} \color[HTML]{000000} 0.174 & {\cellcolor[HTML]{AACFE5}} \color[HTML]{000000} 0.191 & {\cellcolor[HTML]{26627F}} \color{white} 0.163 & {\cellcolor[HTML]{11181C}} \color{white} 0.048 & {\cellcolor[HTML]{180C0A}} \color{white} 1.074 \\
ps \cite{3DS2A} & {\cellcolor[HTML]{0E59A2}} \color[HTML]{F1F1F1} 0.334 & {\cellcolor[HTML]{1C6BB0}} \color[HTML]{F1F1F1} 0.366 & {\cellcolor[HTML]{25607C}} \color{white} 0.161 & {\cellcolor[HTML]{112630}} \color{white} 0.089 & {\cellcolor[HTML]{371000}} \color{white} 0.695 \\
dsgt-bm25-submission \cite{DS@GT} & {\cellcolor[HTML]{3484BF}} \color[HTML]{F1F1F1} 0.283 & {\cellcolor[HTML]{4090C5}} \color[HTML]{F1F1F1} 0.309 & {\cellcolor[HTML]{25607C}} \color{white} 0.159 & {\cellcolor[HTML]{101F26}} \color{white} 0.071 & {\cellcolor[HTML]{4A1502}} \color{white} 0.540 \\
seupd2425-sard-minAnaPhr300 \cite{SARD} & {\cellcolor[HTML]{2575B7}} \color[HTML]{F1F1F1} 0.300 & {\cellcolor[HTML]{3484BF}} \color[HTML]{F1F1F1} 0.328 & {\cellcolor[HTML]{245E7A}} \color{white} 0.157 & {\cellcolor[HTML]{112028}} \color{white} 0.074 & {\cellcolor[HTML]{451401}} \color{white} 0.579 \\
ci \cite{3DS2A} & {\cellcolor[HTML]{2373B6}} \color[HTML]{F1F1F1} 0.302 & {\cellcolor[HTML]{3383BE}} \color[HTML]{F1F1F1} 0.330 & {\cellcolor[HTML]{245D78}} \color{white} 0.155 & {\cellcolor[HTML]{101F26}} \color{white} 0.071 & {\cellcolor[HTML]{3C1101}} \color{white} 0.645 \\
run4 \cite{RACOON} & {\cellcolor[HTML]{1865AC}} \color[HTML]{F1F1F1} 0.320 & {\cellcolor[HTML]{2676B8}} \color[HTML]{F1F1F1} 0.349 & {\cellcolor[HTML]{245D78}} \color{white} 0.154 & {\cellcolor[HTML]{112028}} \color{white} 0.074 & {\cellcolor[HTML]{371000}} \color{white} 0.693 \\
ci-prf-ps \cite{3DS2A} & {\cellcolor[HTML]{2676B8}} \color[HTML]{F1F1F1} 0.299 & {\cellcolor[HTML]{3686C0}} \color[HTML]{F1F1F1} 0.326 & {\cellcolor[HTML]{235B75}} \color{white} 0.153 & {\cellcolor[HTML]{101D25}} \color{white} 0.067 & {\cellcolor[HTML]{3E1201}} \color{white} 0.642 \\
seupd2425-rand-DL-run \cite{RAND} & {\cellcolor[HTML]{B2D2E8}} \color[HTML]{000000} 0.171 & {\cellcolor[HTML]{AED1E7}} \color[HTML]{000000} 0.186 & {\cellcolor[HTML]{225973}} \color{white} 0.149 & {\cellcolor[HTML]{111317}} \color{white} 0.035 & {\cellcolor[HTML]{180C0A}} \color{white} 1.023 \\
ps-prf \cite{3DS2A} & {\cellcolor[HTML]{2B7BBA}} \color[HTML]{F1F1F1} 0.294 & {\cellcolor[HTML]{3A8AC2}} \color[HTML]{F1F1F1} 0.319 & {\cellcolor[HTML]{21556E}} \color{white} 0.144 & {\cellcolor[HTML]{11191E}} \color{white} 0.054 & {\cellcolor[HTML]{3B1100}} \color{white} 0.660 \\
searchill\_YYYY-MM [searchill] & {\cellcolor[HTML]{4B98CA}} \color[HTML]{F1F1F1} 0.258 & {\cellcolor[HTML]{5BA3D0}} \color[HTML]{F1F1F1} 0.277 & {\cellcolor[HTML]{1E4E65}} \color{white} 0.132 & {\cellcolor[HTML]{111317}} \color{white} 0.033 & {\cellcolor[HTML]{8A3F2A}} \color{white} 0.166 \\
bm25-baseline+traditional [air5] & {\cellcolor[HTML]{C7DCEF}} \color[HTML]{000000} 0.149 & {\cellcolor[HTML]{C2D9EE}} \color[HTML]{000000} 0.160 & {\cellcolor[HTML]{1D4B61}} \color{white} 0.127 & {\cellcolor[HTML]{140F10}} \color{white} 0.015 & {\cellcolor[HTML]{1E0B04}} \color{white} 0.934 \\
clef-webbusters-traditional [web-busters] & {\cellcolor[HTML]{519CCC}} \color[HTML]{F1F1F1} 0.253 & {\cellcolor[HTML]{61A7D2}} \color[HTML]{F1F1F1} 0.269 & {\cellcolor[HTML]{194153}} \color{white} 0.112 & {\cellcolor[HTML]{160E0D}} \color{white} 0.008 & {\cellcolor[HTML]{112028}} \color{white} 4.657 \\
mein-cleveres-system \cite{CIR} & {\cellcolor[HTML]{4E9ACB}} \color[HTML]{F1F1F1} 0.255 & {\cellcolor[HTML]{5FA6D1}} \color[HTML]{F1F1F1} 0.271 & {\cellcolor[HTML]{194153}} \color{white} 0.111 & {\cellcolor[HTML]{160E0D}} \color{white} 0.008 & {\cellcolor[HTML]{A65948}} \color{white} 0.000 \\
query\_expansion\_time\_dependence \cite{CIR} & {\cellcolor[HTML]{4D99CA}} \color[HTML]{F1F1F1} 0.256 & {\cellcolor[HTML]{5FA6D1}} \color[HTML]{F1F1F1} 0.272 & {\cellcolor[HTML]{194153}} \color{white} 0.111 & {\cellcolor[HTML]{170D0B}} \color{white} 0.007 & {\cellcolor[HTML]{8C402C}} \color{white} 0.147 \\
bm25\_boosted\_temporal\_categories \cite{CIR} & {\cellcolor[HTML]{89BEDC}} \color[HTML]{000000} 0.206 & {\cellcolor[HTML]{97C6DF}} \color[HTML]{000000} 0.213 & {\cellcolor[HTML]{11242E}} \color{white} 0.064 & {\cellcolor[HTML]{391100}} \color{white} -0.035 & {\cellcolor[HTML]{300F00}} \color{white} 0.760 \\
luyuHF [air5] & {\cellcolor[HTML]{CCDFF1}} \color[HTML]{000000} 0.141 & {\cellcolor[HTML]{CBDEF1}} \color[HTML]{000000} 0.145 & {\cellcolor[HTML]{101F26}} \color{white} 0.053 & {\cellcolor[HTML]{340F00}} \color{white} -0.030 & {\cellcolor[HTML]{270D01}} \color{white} 0.842 \\
dense-baseline+multilingual [air5] & {\cellcolor[HTML]{F7FBFF}} \color[HTML]{000000} 0.073 & {\cellcolor[HTML]{F7FBFF}} \color[HTML]{000000} 0.054 & {\cellcolor[HTML]{FFADAD}} \color{black} -1.129 & {\cellcolor[HTML]{FFADAD}} \color{black} -0.165 & {\cellcolor[HTML]{330F00}} \color{white} 0.727 \\
\bottomrule
\end{tabular}
}
\end{table}

\begin{table}[!htb]
\caption{Change results for the SciRetrieval task in comparison to the snapshot 2024-11. The results are sorted by RI. Darker colors indicate less change or higher effectiveness.}
\label{tab:sci-results-change}
\resizebox{\textwidth}{!}{%
\begin{tabular}{lc@{\hskip 5pt}c@{\hskip 5pt}c@{\hskip 5pt}c@{\hskip 5pt}c}
\toprule
Approach & ARP & AARP & RI & DRI & ER \\
\midrule
fusion-with-core \cite{OWS} & {\cellcolor[HTML]{D0E1F2}} \color[HTML]{000000} 0.059 & {\cellcolor[HTML]{B7D4EA}} \color[HTML]{000000} 0.060 & {\cellcolor[HTML]{9EB0FF}} \color{black} 0.029 & {\cellcolor[HTML]{9EB0FF}} \color{black} 0.108 & {\cellcolor[HTML]{111A20}} \color{white} 1.523 \\
monot5-in-core \cite{OWS} & {\cellcolor[HTML]{E2EDF8}} \color[HTML]{000000} 0.031 & {\cellcolor[HTML]{DBE9F6}} \color[HTML]{000000} 0.029 & {\cellcolor[HTML]{180C0A}} \color{white} -0.120 & {\cellcolor[HTML]{132D3A}} \color{white} 0.025 & {\cellcolor[HTML]{111317}} \color{white} 1.312 \\
ows-bm25 \cite{OWS} & {\cellcolor[HTML]{2171B5}} \color[HTML]{F1F1F1} 0.218 & {\cellcolor[HTML]{083C7D}} \color[HTML]{F1F1F1} 0.191 & {\cellcolor[HTML]{180C0A}} \color{white} -0.339 & {\cellcolor[HTML]{210B03}} \color{white} -0.058 & {\cellcolor[HTML]{112028}} \color{white} 1.673 \\
ows-cluster-boosting \cite{OWS} & {\cellcolor[HTML]{1F6EB3}} \color[HTML]{F1F1F1} 0.222 & {\cellcolor[HTML]{083A7A}} \color[HTML]{F1F1F1} 0.193 & {\cellcolor[HTML]{180C0A}} \color{white} -0.363 & {\cellcolor[HTML]{240C02}} \color{white} -0.080 & {\cellcolor[HTML]{112630}} \color{white} 1.816 \\
rm3-on-qrel-boost \cite{OWS} & {\cellcolor[HTML]{0A539E}} \color[HTML]{F1F1F1} 0.253 & {\cellcolor[HTML]{08306B}} \color[HTML]{F1F1F1} 0.201 & {\cellcolor[HTML]{180C0A}} \color{white} -0.680 & {\cellcolor[HTML]{591C07}} \color{white} -0.351 & {\cellcolor[HTML]{9EB0FF}} \color{black} 5.257 \\
A3-TFIDF-TestAbs [sambs] & {\cellcolor[HTML]{77B5D9}} \color[HTML]{000000} 0.137 & {\cellcolor[HTML]{69ADD5}} \color[HTML]{F1F1F1} 0.102 & {\cellcolor[HTML]{180C0A}} \color{white} -1.028 & {\cellcolor[HTML]{4D1602}} \color{white} -0.294 & {\cellcolor[HTML]{250C01}} \color{white} 0.585 \\
A4-TFIDF-TestFull [sambs] & {\cellcolor[HTML]{7CB7DA}} \color[HTML]{000000} 0.134 & {\cellcolor[HTML]{6CAED6}} \color[HTML]{F1F1F1} 0.100 & {\cellcolor[HTML]{180C0A}} \color{white} -1.028 & {\cellcolor[HTML]{4B1602}} \color{white} -0.287 & {\cellcolor[HTML]{240C02}} \color{white} 0.619 \\
SciBERT\_LongEVAL [long-eval-sci-group-5] & {\cellcolor[HTML]{D0E1F2}} \color[HTML]{000000} 0.059 & {\cellcolor[HTML]{CDE0F1}} \color[HTML]{000000} 0.043 & {\cellcolor[HTML]{180C0A}} \color{white} -1.132 & {\cellcolor[HTML]{2D0E00}} \color{white} -0.137 & {\cellcolor[HTML]{170D0B}} \color{white} 1.057 \\
bm25+reranker+weighted [tf-idk] & {\cellcolor[HTML]{4997C9}} \color[HTML]{F1F1F1} 0.176 & {\cellcolor[HTML]{4695C8}} \color[HTML]{F1F1F1} 0.123 & {\cellcolor[HTML]{180C0A}} \color{white} -1.482 & {\cellcolor[HTML]{843924}} \color{white} -0.494 & {\cellcolor[HTML]{3B1100}} \color{white} -0.048 \\
B3-SBERT-V1-TestAbs [sambs] & {\cellcolor[HTML]{CFE1F2}} \color[HTML]{000000} 0.060 & {\cellcolor[HTML]{CEE0F2}} \color[HTML]{000000} 0.042 & {\cellcolor[HTML]{180C0A}} \color{white} -1.517 & {\cellcolor[HTML]{340F00}} \color{white} -0.170 & {\cellcolor[HTML]{170D0B}} \color{white} 1.051 \\
bm25+reranker [tf-idk] & {\cellcolor[HTML]{4191C6}} \color[HTML]{F1F1F1} 0.183 & {\cellcolor[HTML]{4191C6}} \color[HTML]{F1F1F1} 0.126 & {\cellcolor[HTML]{180C0A}} \color{white} -1.663 & {\cellcolor[HTML]{944834}} \color{white} -0.553 & {\cellcolor[HTML]{3F1201}} \color{white} -0.165 \\
BM25\_k1\_0p95\_b\_0p75\_stop\_stem\_fullText [academy-retrievals] & {\cellcolor[HTML]{08316D}} \color[HTML]{F1F1F1} 0.290 & {\cellcolor[HTML]{083573}} \color[HTML]{F1F1F1} 0.197 & {\cellcolor[HTML]{180C0A}} \color{white} -1.817 & {\cellcolor[HTML]{FBA9A8}} \color{black} -0.920 & {\cellcolor[HTML]{FFADAD}} \color{black} -3.696 \\
BM25\_k1\_0p2\_b\_0p8\_stop\_stem\_fullText [academy-retrievals] & {\cellcolor[HTML]{08306B}} \color[HTML]{F1F1F1} 0.291 & {\cellcolor[HTML]{083674}} \color[HTML]{F1F1F1} 0.196 & {\cellcolor[HTML]{180C0A}} \color{white} -1.862 & {\cellcolor[HTML]{FFADAD}} \color{black} -0.935 & {\cellcolor[HTML]{F9A7A6}} \color{black} -3.569 \\
BM25\_k1\_1p0\_b\_0p7\_stop\_stem\_fullText [academy-retrievals] & {\cellcolor[HTML]{08306B}} \color[HTML]{F1F1F1} 0.292 & {\cellcolor[HTML]{083573}} \color[HTML]{F1F1F1} 0.197 & {\cellcolor[HTML]{180C0A}} \color{white} -1.865 & {\cellcolor[HTML]{FFADAD}} \color{black} -0.939 & {\cellcolor[HTML]{FDABAB}} \color{black} -3.628 \\
B4-SBERT-V2-MPNet-TestFull [sambs] & {\cellcolor[HTML]{CADDF0}} \color[HTML]{000000} 0.068 & {\cellcolor[HTML]{CADEF0}} \color[HTML]{000000} 0.046 & {\cellcolor[HTML]{180C0A}} \color{white} -1.867 & {\cellcolor[HTML]{3C1101}} \color{white} -0.218 & {\cellcolor[HTML]{190C09}} \color{white} 0.978 \\
B3-SBERT-V2-MPNet-TestAbs [sambs] & {\cellcolor[HTML]{CADEF0}} \color[HTML]{000000} 0.067 & {\cellcolor[HTML]{CBDEF1}} \color[HTML]{000000} 0.045 & {\cellcolor[HTML]{180C0A}} \color{white} -1.892 & {\cellcolor[HTML]{3C1101}} \color{white} -0.215 & {\cellcolor[HTML]{190C09}} \color{white} 0.983 \\
BM25\_ColBERTv2\_ReRanker\_v1 [academy-retrievals] & {\cellcolor[HTML]{4896C8}} \color[HTML]{F1F1F1} 0.177 & {\cellcolor[HTML]{4D99CA}} \color[HTML]{F1F1F1} 0.119 & {\cellcolor[HTML]{180C0A}} \color{white} -1.916 & {\cellcolor[HTML]{9A4D3B}} \color{white} -0.575 & {\cellcolor[HTML]{3B1100}} \color{white} -0.031 \\
qrel-boost-core \cite{OWS} & {\cellcolor[HTML]{1966AD}} \color[HTML]{F1F1F1} 0.231 & {\cellcolor[HTML]{2777B8}} \color[HTML]{F1F1F1} 0.146 & {\cellcolor[HTML]{180C0A}} \color{white} -2.789 & {\cellcolor[HTML]{F2A19F}} \color{black} -0.888 & {\cellcolor[HTML]{5D1E09}} \color{white} -0.800 \\
query-intent-fusion \cite{OWS} & {\cellcolor[HTML]{1966AD}} \color[HTML]{F1F1F1} 0.231 & {\cellcolor[HTML]{2777B8}} \color[HTML]{F1F1F1} 0.146 & {\cellcolor[HTML]{180C0A}} \color{white} -2.789 & {\cellcolor[HTML]{F2A19F}} \color{black} -0.888 & {\cellcolor[HTML]{5D1E09}} \color{white} -0.800 \\
\bottomrule
\end{tabular}
}
\end{table}

\section{Discussion}
This year, more teams than ever registered for the LongEval Lab. In total, 56 teams registered at CLEF, and 55 teams also registered at TIRA to submit an approach. Of those, 19 teams submitted at least one approach, which is the highest number of teams to date. 68 approaches were submitted, almost twice as many as last year, but still six submissions fewer compared to the first iteration in 2023. Twelve notebook papers were submitted, while 14 were received in 2023 and nine in 2024. For the first time, LongEval incorporated a new retrieval task: SciRetrieval. As expected, this task received fewer submissions compared to the established WebRetrieval task. Besides being new, this might also be because fewer snapshots are available so far, an important factor for temporal systems. Nevertheless, the task received many interesting approaches and was generally well received.

The new task is a big step, as longitudinal evaluation is not primarily a web search-related task but is imminent in very different retrieval settings. Therefore, a task designed for a new and disjoint dataset from a different retrieval domain is a very fruitful addition to LongEval. SciRetrieval originates from the CORE platform, one of the largest scientific retrieval systems worldwide, with currently 441 million publicly available research papers indexed. This dataset is especially interesting because it covers the unique dynamics of a very different retrieval setting that can be described as academic search. While some other labs also used scientific papers to create a dataset or retrieval task, like the TREC domain-specific cross-language IR task~\cite{DBLP:conf/clef/KluckG00} or the LiLAS lab at CLEF~\cite{DBLP:conf/clef/SchaerBCWST21a}, none of them included a developing test collection with different time frames and snapshots. The only academic search dataset that allows for studying longitudinal effects to some extent is TREC-COVID~\cite{DBLP:journals/sigir/VoorheesABDHLRS20}.

Academic search and our new collaboration with CORE enables the community to create a test collection and shared task that includes new interaction paradigms, like faceted search or queries using classic patterns like block strategies or specific scientific search stratagems such as ``Footnote Chasing'', ``Citation Searching'', `Keyword Searching'', ``Author Searching'' and ``Journal Run'' as described in the seminal paper of Marcia Bates in 1990~\cite{DBLP:journals/ipm/Bates90}. These stratagems enable going beyond simple ad-hoc search settings and are a step toward evaluating interactive retrieval settings and search sessions. 

Both tasks follow the same two main objectives of the lab: To explore how the retrieval effectiveness drops over time and to develop systems resilient to that. Regarding the drop in effectiveness, this year the focus shifted towards measuring changes. In the first iteration of the lab in 2023, the effectiveness improved on average, while in 2024 it actually dropped~\cite{longevaloverview2023,longevalCLEF2024overview}. Overall, this year the effectiveness is only slightly decreasing, although last years datasets were also based on the snapshots 2023-06 and 2023-08. These observations highlight the importance of the reference point to which the results are compared.

Specifically, as shown in Table~\ref{tab:web-results}, a clear decline in nDCG@10 can be observed across systems from snapshot 2023-03 to 2023-08 in most approaches. This aligns with the document overlap matrix in Figure~\ref{fig:webretrieval_docs_overlap}, which illustrates that later snapshots contain increasingly different content compared to earlier ones (i.e. starting from 2023-06). This decreasing overlap suggests a shifting document distribution over time, which presents a major challenge for static systems. This highlights again the importance of designing retrieval models that remain effective across temporally distant test collections.

It was observed again how the most effective approaches do not necessarily align with the approaches with the fewest change. This motivates us to continue this objective and research more complete definitions of temporal robustness and persistence. Investigations in this regard were also adopted from some teams as complementary evaluations in their notebook papers. In contrast regarding the second objective, improving the retrieval effectiveness remained the main motivation for all teams and no team specifically focused on developing a system with an especially stable effectiveness. 

The submitted approaches increasingly rely on past snapshots and approaches that directly use previous qrels show to be highly effective especially in the WebRetrieval task where a long history of snapshots is available. While many interesting relevance signals are developed from this data, no team submitted a retrieval model that was directly trained or fine-tuned on prior snapshots. This remains as an interesting direction for future work.

\section{Conclusion}
We have given an overview of the LongEval 2025 Lab, describing briefly the data provided for the two tasks, the participant submissions, and the carried out evaluation. Compared to the previous years, participants have designed a variety of solutions to handle the temporal aspects of the given document collections.
The approaches adopted by the participants to LongEval 2025 highlight the fast-paced evolution of IR systems. While established retrieval methods such as BM25 and Lucene remain main building blocks, we notice a shift toward integrating advanced methods like chunk-based indexing, neural reranking, and query expansion powered by large language models (LLMs). Some teams prioritized computational efficiency and broad accessibility, optimizing classical IR frameworks for use on standard hardware with multithreaded processing. Others combined neural solutions, incorporating statistical analyses and topic modeling to more effectively address the temporal and semantic challenges presented by contemporary queries.
Retrieval effectiveness was measured using nDCG and nDCG@10, looking at how well systems ranked relevant documents in the top results for the WebRetrieval task. Only the best or latest versions of similar submissions were considered in our analysis. The best performing approaches to the WebRetrieval task remained stable over time, with Team RISE leading in later snapshots. For the SciRetrieval task, leading approaches varied by snapshot, but system rankings were generally consistent across evaluation measures.

We also analysed the changes in retrieval effectiveness changes over time to identify which systems remain robust as the collections to search in evolve. We used Relative Improvement (RI), Delta Relative Improvement (DRI), and Effect Ratio (ER) metrics, which we compared to the BM25 baseline. We claim that these measures help isolate the effect of system changes from general shifts in the search environment. Results show that while many systems improved in the short-term setting, only a few maintained or even enhanced their effectiveness long-term, with most changes being very small. For the Sciretrieval task, the persistence and improvement of system effectiveness varied depending on the metric used, with some systems showing high robustness (ER close to 1), though the different measures did not always agree on which systems performed best over time.

\begin{credits}
\subsubsection{\ackname}
This work is supported by the ANR Kodicare bi-lateral project, grant ANR-19-CE23-0029 of the French Agence Nationale de la Recherche, the Austrian Science Fund (FWF, grant I4471-N), the UKRI/EPSRC Turing AI Fellowship to Maria Liakata (grant no. EP/V030302/1), the Ministry of Education, Youth and Sports of the Czech Republic, Project No. LM2023062 LINDAT/CLARIAH-CZ, and the German Research Foundation (DFG) through project grant No. 407518790.
This work has been using services provided by the LINDAT/CLARIAH-CZ Research Infrastructure (https://lindat.cz), supported by the Ministry of Education, Youth and Sports of the Czech Republic (Project No. LM2023062).

\subsubsection{\discintname}
The authors have no competing interests to declare that are relevant to the content of this article.
\end{credits}

\bibliographystyle{splncs04}
\bibliography{Longeval}

\newpage
\appendix

\end{document}